\documentclass[10pt,english,french,babel]{article}
\usepackage[T1]{fontenc}
\usepackage[latin1]{inputenc}
\usepackage{babel}
\usepackage{graphics}

\makeatletter

\providecommand{\LyX}{L\kern-.1667em\lower.25em\hbox{Y}\kern-.125emX\@}
\let\SF@@footnote\footnote
\def\footnote{\ifx\protect\@typeset@protect
    \expandafter\SF@@footnote
  \else
    \expandafter\SF@gobble@opt
  \fi
}
\expandafter\def\csname SF@gobble@opt \endcsname{\@ifnextchar[
  \SF@gobble@twobracket
  \@gobble
}
\edef\SF@gobble@opt{\noexpand\protect
  \expandafter\noexpand\csname SF@gobble@opt \endcsname}
\def\SF@gobble@twobracket[#1]#2{}

\usepackage[T1]{fontenc}
\usepackage{graphics}
\usepackage{color}


\begin{document}
\setlength{\textwidth}{27pc}
\selectlanguage{english}

\title{Classification of phase transitions and  ensemble inequivalence, in systems with long range interactions\\
 }

\author{F. BOUCHET{\small \( ^{1,2} \)} and J. BARR\'{E} {\small \( ^{2,1,3} \)}\\
\\
 {\small \( ^{1} \)Dipartimento di Energetica {}``Sergio Stecco''.
Via S. Marta, 3. 50137 Firenze. Italia.}\\
 {\small \( ^{2} \)\'Ecole Normale Supérieure de Lyon, 46 all\'ee
   d'Italie, 69364 Lyon Cedex 07, France.}\\
 {\small $^{3}$ Theoretical Division, Los Alamos National Laboratory.
  Los Alamos, NM87545, USA.} 
}

\maketitle
\begin{abstract}
Systems with long range interactions in general are not additive, which
can lead to an inequivalence of the microcanonical and canonical ensembles.
The microcanonical ensemble may show richer behavior than the canonical one, 
including negative specific heats and other non-common behaviors.
We propose a classification of microcanonical phase transitions, of
their link to canonical ones, and of the possible situations of ensemble
inequivalence. We discuss previously observed phase transitions and
inequivalence in self-gravitating, two-dimensional fluid dynamics
and non-neutral plasmas. We note a number of generic situations that
have not yet been observed in such systems.
\end{abstract}
\vskip 1cm 

\hskip -0.5cm 
\noindent
Author contact: Freddy Bouchet \\
e-mail: Freddy.Bouchet@ens-lyon.fr \\
Phone: (33) 4 72 72 81 48 \\
Address: {\small Laboratoire de physique, Ecole Normale Supérieure de Lyon\\
46 all\'ee d'Italie, 69364 Lyon Cedex 07, France. \par{}}{\small \par}
{\small \nonumber{}\par{}}{\small \par}
\vskip .5cm

\noindent
\underline{suggested running head}:\\
Ensemble inequivalence and long range interactions: a classification.
\newpage

\section{Introduction\label{sec_Introduction}}

In a large number of physical systems, any single particle experiences
a force which is dominated by interactions with far away particles.
For instance, in a system with algebraic decay of the inter-particle
potential \( V\left( r\right) \sim _{r\rightarrow \infty }1/r^{\alpha } \),
when \( \alpha  \) is less than the dimension of the system, the
interaction is long range (such interactions are sometimes called ``non-integrable''). Such long range
interacting systems are
not-additive, as the interaction of any macroscopic part of the system with the
whole is not negligible with respect to the internal energy of the given part. 

The main physical examples of non-additive, long range interacting systems
are: astrophysical self-gravitating
systems \cite{Chavanis_Houches,De Vega,Lynden_Bell,millerbruce,Padmanabhan,
Stahl_Kiessling_Schindler,Votyakov}, two-dimensional or 
geophysical fluid dynamics
\cite{Chavanis_Sommeria00,miller,rob som 91}, 
and certain plasma physics models \cite{firpo,Kiessling_Neukirch}.
Spin systems \cite{barre,ispolatov} and toy models \cite{Antoni2}
with long range interactions have also been widely studied. 

As a consequence of the lack of additivity, peculiar thermodynamic
behaviors are likely to be observed in such Hamiltonian systems. For
instance, the usual proof of the validity of the canonical ensemble,
for a system in contact with a thermostat, or for a part of a bigger
isolated system, uses explicitly this additivity property. Hertel and
Thirring provided a toy model, mimicking self-gravitating
dynamics~\cite{hertel}, which displays inequivalence between canonical
and microcanonical solutions, with negative specific heat regions in
the microcanonical ensemble. Negative specific heat and ensemble
inequivalence, previously known to astrophysicists, were then found in
various fields: plasma physics~\cite{Kiessling_Neukirch,Smith_ONeil},
2D fluid dynamics~\cite{Caglioti} and geophysical fluid
dynamics~\cite{ellis}. These examples show that new
  types of phase transitions are found in long range interacting
  systems.

A natural question then arises: do we know all possible behaviors
stemming from long range interactions, and, if not, what are the
possible phenomenologies? The aim of this article is to answer the
question by providing a \emph{classification} of all microcanonical
and canonical phase transitions, in long range interacting systems,
with emphasis on situations of ensemble inequivalence. Although we
restrict here to long range interacting systems, we want to stress that
these systems are not the only ones for which ensemble inequivalence
may occur~\cite{Les_Houches}.
Let us note also that the deep link between ensemble inequivalence and
dynamical non-linear stability issues has been recently recognized
\cite{Ellis_Stabilite}, which provides a further
incentive for this study. \\

In order to formalize the problem, we first argue, in section
\ref{sec_Mean_Field}, that the mean field approach is exact, in the
limit of a large number of particles, for most systems with
long range interactions. In this mean field context, the
microcanonical equilibrium is defined by the maximization, with
respect to an order parameter, of an entropy with an energy
constraint. The canonical equilibrium is then given by the
minimization of the associated free energy. The study of phase
transitions is thus reduced to the study of the singularities of these
two variational problems, while the study of ensemble inequivalence is reduced
to comparing the equilibrium states in both ensembles. 

We propose a classification of all singularities, convexity changes, 
and convexification properties of entropy functions,
independently of the underlying physical problem; it uses the tools of
singularity theory, along the lines of the works of Varchenko
\cite{Varchenko} and Aicardi \cite{Aicardi2001}, on the classification
of phase transitions for binary mixtures, in classical thermodynamics.
We thus classify all microcanonical and canonical phase transitions,
their mutual link, and all situations of ensemble inequivalence.

In section \ref{sec_Classification}, we  carry on this program
  when one internal (the energy for instance) and one external
  parameter are varied ; we then review the transitions already found
in the (known to us) literature, by studying actual $N$-body systems.
We will see that the classification identifies many new possibilities
besides the well-known negative specific heat regions. In section
\ref{sec_Classification_Symetrie}, we give a similar classification
and comparison with the existing literature for systems with a parity
symmetry.

In all this work, we study the possible inequivalence of ensembles,
when only one dynamical constraint is taken into account. This is
valid for inequivalence between the microcanonical and the canonical
ensembles, or between the canonical and grand canonical ensembles, for
instance. We briefly discuss the generalization to several
constraints in the conclusion.

\section{Mean field statistical mechanics and ensemble inequivalence\label{sec_Mean_Field}}

\subsection{Microcanonical and canonical equilibrium states\label{sec_Grd_Deviations}}

A model will be said to have long range interactions when any single
particle experiences a force for which a macroscopic number of
particles contribute, and such that the contribution of closest
particles is negligible when the number of particles \( N \) goes to
infinity. The statistical equilibrium states of such systems are
generically described by mean-field variational problems. We first
give some heuristic justification of this statement, before referring
to rigorous proofs for specific models.

The energy of such systems can be approximated, in the large $N$
limit, by the energy of a coarse-grained variable $m$ which may be a
scalar, a vector or a field ($H\sim _{N\rightarrow \infty
  }Nh\left( m\right)$). The description of equilibrium structures then
amounts to compute the probability $P(m)$ of this coarse-grained
field. This probability is characterized by an entropy function or
functional in the large $N$ limit: $ \log \left( P \left( m\right)
\right) \sim _{N\rightarrow \infty }Ns\left( m\right)$. As a result,
the microcanonical equilibrium state \( m_{m} \) and entropy of the
system \( S(E) \), for a given energy \( E \), are the solutions of
the maximization of the entropy  function (functional) \( s \) 
with the energy constraint (this general approach is explained
in~\cite{ellis}):
\begin{equation}
\label{varproblem}
S(E)=\sup _{m}\left\{ s(m)\, \, |\, \, h(m)=E\right\} =s(m_{m}).
\end{equation}  
The points where the supremum is reached defines \( m_{m}(E) \), the
microcanonical equilibrium state at energy \( E \). We suppose that it
exists\footnote{the existence has to be proved for each case, using
  properties of the entropy functional \( s(m) \), which is usually
  strictly concave.}; however it is not necessarily unique.
 
With similar arguments, the canonical\footnote{in systems with long
  range interaction, the canonical ensemble does not describe the
  fluctuations of a small subsystem. However it can describe
  fluctuation of the whole system coupled to a thermostat with
  vanishingly small coupling.} equilibrium \( m_{c} \), at inverse
temperature \( \beta \), is given by the minimization of the free
energy functional:
\begin{equation}
\label{varproblemcano1}
F(\beta )=\inf _{m}\left\{ -s(m)+\beta h(m)\right\} =-s(m_{c})+\beta h(m_{c})
\end{equation}
 We note that the usual free energy is \( F\left( \beta \right) /\beta  \).
Nevertheless, for sake of simplicity, we will always call \( F \)
the free energy
\footnote{
 For negative temperature states, for instance in two dimensional
turbulence, the functional $-s(m)+\beta h(m)$ must still be minimized,
whereas the usual free energy should be maximized. Such a notation
thus simplifies the discussion.}. 
The points where the infimum is reached defines \( m_{c}(\beta ) \),
the canonical equilibrium state at inverse temperature \( \beta  \);
we suppose that it exists, but as in the microcanonical case, it is
not necessarily unique. In a canonical context, this reduction to a
variational problem was already rigorously described
  for gravitating fermions in~\cite{Hertel71}, and in a more general
  setting in~\cite{Messer82}.

Such a mean field description has been proposed for the point-vortex
system \cite{Joyce_Montgommery72}, two dimensional incompressible
flows \cite{miller,robert}, Quasi-geostrophic flows, rotating
Shallow-Water model \cite{Chavanis_Sommeria00} \footnote{The main
  peculiarity of models of two dimensional or quasi-geostrophic flows
  is the existence of an infinite number of conserved quantities
  (Casimirs), due to the continuous nature of the dynamics, and
  similar to the conserved quantity of a Vlasov dynamics.},
self-gravitating systems \cite{Chavanis_Houches,Lynden_Bell,
  Padmanabhan,Stahl_Kiessling_Schindler}, plasma physics
\cite{firpo,Kiessling_Neukirch}, spin systems
\cite{Antoni2,barre,ispolatov}. Some rigorous large deviations
 results, confirming this mean field description, have been
rigorously obtained for a large class of long range interacting
systems: two-dimensional or quasi-geostrophic models
\cite{ellis,michel,Robert98}, the point vortex model
\cite{Caglioti,Kiessling_Lebowitz}, spin systems \cite{barre}.

From now on, we call microcanonically stable,
  metastable and unstable state respectively a global maximum of
  problem (\ref{varproblem}) (that is a microcanonical equilibrium
  state), a local maximum of (\ref{varproblem})
  which is not global, and a local minimum or a saddle point of
  (\ref{varproblem}). We define similarly canonically stable,
  metastable and unstable states with problem
  (\ref{varproblemcano1}).  In the following we will study generic
properties of the two dual variational problems (\ref{varproblem}) and
(\ref{varproblemcano1}), independently of any specific system. The
Lagrange multiplier result insures that, for a given energy $E$, a
critical point of (\ref{varproblem}) is a critical point of
(\ref{varproblemcano1}) for some value of $\beta$, verifying \(
-ds+\beta dh=0 \) \footnote{By definition, the microcanonical inverse temperature $\beta$ is the Lagrange parameter $\beta$ corresponding to the entropy maximum (it may be not uniquely defined, for instance at a first order microcanonical transition point). When  $S$ is differentiable, we have $\beta=s'(E)$. On the contrary, in the canonical ensemble, $\beta$ is a parameter. If the two ensembles are equivalent at the energy $E$, then the canonical calculations with
inverse temperature $\beta=s'(E)$ will yield the same equilibrium state and the energy $E$. When no confusion is possible, $\beta$ will either refer to the microcanonical or to the canonical inverse temperature}. However the stability of this critical point
(stable, metastable, unstable) may differ for the
two variational problems. We will say that the microcanonical and
canonical ensembles are equivalent at an energy $E$ if it exists
$\beta$ such that $m_m (E)$=$m_c (\beta)$ (with $m_c(\beta)$
  canonically stable).
In the opposite case, we will say that the two ensembles are not
equivalent. The problem of ensemble equivalence thus reduces to the
study of the solutions of the two general variational
problems~(\ref{varproblem}) and~(\ref{varproblemcano1}). In the
following paragraph, we recall the results linking ensemble
equivalence and the concavity of $S(E)$.

\subsection{Characterization of ensemble equivalence \label{sec_caracterisation}}

It is classically known that ensemble equivalence is related to the
concavity of the entropy \( S\left( E\right) \) (for instance authors
of \cite{Kiessling_Lebowitz} use this property as a definition). If
the entropy is twice differentiable and non concave at some point,
both ensembles are obviously not equivalent, since the specific heat is
always positive in the canonical
ensemble \footnote{The specific heat is defined in the
    canonical ensemble as $C_v^{\mbox{cano}}=-\beta^2 d^2F/d\beta^2$,
    and as $C_v^{\mbox{micro}}=-\beta^2/ (d^2S/dE^2)$ in the
    microcanonical ensemble.}. More generally, ensemble inequivalence
is completely characterized by the concavity of \( S\left( E\right)
\).  The concave envelope of the $S(E)$ curve is defined as the
boundary of the intersection of all closed half-spaces which contains
the $S(E)$ curve. One can prove the three following points by convex
analysis~\cite{ellis}:
\begin{enumerate}
\item a canonical equilibrium state at inverse temperature \( \beta  \)
is always a microcanonical equilibrium state for some energy \( E \). 
\item a microcanonical equilibrium state of energy \( E \) is also a canonical
equilibrium state, for some inverse temperature \( \beta  \), if
and only if the function \( S \) coincides, at the point \( E \),
with its concave envelope. 
\item the second point can be refined: we consider an energy \( E \) at
which \( S \) is regular (at least twice differentiable); then a
microcanonical equilibrium state of energy \( E \) is a canonical
stable or metastable state, for some inverse temperature \( \beta  \),
if, and only if, the function \( S \) is locally concave around the
point \( E \) (that is locally under its tangent). 
\end{enumerate}
A proof of points 1 and 2, in the context of long range interacting
systems, from the variational problems (\ref{varproblem}) and
(\ref{varproblemcano1}), is given in~\cite{ellis}. Point 3 is a
direct generalization.  \footnote{Here is a sketch of a possible
  proof: a canonical state \( m_{c} \), with energy \( E=h\left(
    m_{c}\right) \), is by definition said to be metastable if it is a
  local, and not a global minimum of the free energy functional (for an
  infinite dimensional \( m \), one has to specify explicitly some
  topology). If the energy is continuous, states close to \( m_{c} \)
  have an energy close to $E$. The application of point 2 to a class
  of states with energy close to $E$ proves 3.}

Let us furthermore argue that the knowledge of the canonical
equilibria \( m_{c}(\beta ) \), for all \( \beta \), is sufficient to
know whether ensemble inequivalence occurs or not. From the knowledge
of all \( m_{c}(\beta ) \), we can compute all canonical equilibrium
energies \( \left\{ h\left( m_{c}(\beta )\right) \right\} \). If this
energy ensemble is the same as the energy range of the system, then
the whole curve \( S(E) \) can be constructed from \( \left\{ \left(
    h\left( m_{c}(\beta )\right) ,s\left( m_{c}(\beta )\right) \right)
\right\} \). This curve is always concave, as a canonical equilibrium
(not necessarily unique) may be associated to each point. Thus there
is no ensemble inequivalence. Conversely, if the canonical energy
range is different from the system energy range, then for these
energies, microcanonical equilibria are not canonical ones, and
ensembles are inequivalent. This discussion also shows that a
microcanonical phase diagram, where the convexity changes and
convexification points or surface are represented, contains all
information necessary to construct the associated canonical phase
diagram (the converse is wrong in general).

The problem of ensemble inequivalence is thus reduced to the study of
the concavity of the function \( S(E) \). More generally,
microcanonical phase transitions then correspond to a lack of
analyticity of the entropy \( S\left( E\right) \), whereas ensemble
inequivalence or canonical phase transitions are characterized by
convexity changes of the entropy \( S\left( E\right) \). We note that in systems with short range interactions, the mechanism of phase separation justifies the Maxwell construction, and explain why entropy functions are generally concave. 
These considerations are model independent, which leads to the central point
of the paper, in section 3 and 4: we have the opportunity to classify
all the possible phase transitions for these systems (or equivalently,
to classify the analyticity breakings, convexification and concavity
properties of $S(E)$); we will make use for this purpose of the tools
of singularity theory~\cite{arnold} (or catastrophe theory~\cite{Thom}).

\section{Classification of long-range interacting systems phase-transitions
and of statistical ensemble inequivalence\label{sec_Classification} }


In this section, we focus on the case where no symmetry property is
assumed for the dependence of the functionals $s$ and $h$ on the coarse-grained variable $m$. We classify all
maximization, convexity and convexification singularities of an
entropy-energy curve. For each situation, we are interested in
singularities which are not removed by any small perturbation of the
functions $s$ and $h$ defining the variational problems
(\ref{varproblem}) and (\ref{varproblemcano1}). Let us suppose that
$s$ and $h$ depend on $N+1$ parameters; heuristically we will say that
a singularity is generic with codimension $n$ if it spans an $N-n$
dimensional hypersurface in the $N+1$~dimensional parameter space.
Thus, if one explores an $n+1$ dimensional hypersurface in that space,
that is if $n+1$ parameters can be varied, a codimension~$n$
singularity will be {\it generically} observed. If we consider the
energy $E$ to be one of these parameters (the internal one), $n$ then
refers to the number of external parameters which has to be varied to
observe {\it generically} codimension $n$ singularities. Varchenko
\cite{Varchenko} classified singularities of the convex envelope of a
finite set of smooth bounded function, up to codimension 2.
Aicardi~\cite{Aicardi2001} completed this classification and applied
it to the classification of phase transitions for binary mixtures.
These works do not however consider maximization singularities.
Brysgalova ~\cite{Brysgalova} classified maximization singularities
for variational problems depending on a parameter. We know no
mathematical results classifying both singularities and
convexification properties of a variational problem.
We now first describe ``codimension -1'' points (see
  below), and classify all codimension 0 and 1 singularities (that is
with none or one external parameter), when only one constraint (the
energy) is considered. We then search in the literature which of these
transitions have been found in actual physical models of interacting
particles.

\subsection{Generic {\bf points on a $S(E)$ curve} \label{sec_Generic}}

We describe here \emph{generic points} of a $S(E)$ curve, that is 
points accessible without tuning the energy. 
 Formally, these would be ``codimension~$-1$'' singularities.
We thus consider the three sources of
singularity we are interested in: maximization singularity, convexity
property and convexification. The corresponding generic points are:

\begin{itemize}
\item \textbf{Maximization singularities and analytic properties}\\
 A generic point with respect to maximization is a point where \( S(E) \)
is analytic. 
\item \textbf{Convexity properties}\\
 A generic point with respect to the convexity properties is a point
where the second derivative of \( S(E) \) is non zero, so that \( S \)
has a definite concavity. This defines two types of generic points
with respect to the convexity properties, the concave and the convex
ones. 
\item \textbf{Properties with respect to convexification}\\
 A generic point of the curve \( S(E) \) may or may not belong to
the concave envelop of \( S \). Thus, there are two types of generic
points with respect to convexification. 
\end{itemize}
We may now combine the different properties to get the different types
of generic points: 

\begin{itemize}
\item A concave point may or may not belong to the concave envelop of
  \( S \).  Consequently, there are two types of concave generic
  points: those who belong to the concave envelop (points \( A \) on
  Fig.~\ref{codim-1}), and those who do not belong to the envelop
  (points \( C \) on Fig.~\ref{codim-1}).
\item A convex point cannot belong to the concave envelop, so that
  there is only one type of such generic points (points \( B \) on
  Fig.~\ref{codim-1}).\\

\end{itemize}
We have now classified the three types of generic points. The microcanonical
and canonical ensembles are equivalent at points of the \( A \) type,
and inequivalent at points of the \( B \) or \( C \) types (\( B \)
or \( C \) define the \emph{inequivalence range}). The
negative specific heat regions correspond to points of the \( B \)
type.
The range of admissible values for \( E \) can be decomposed into several 
intervals, in the
interior of which \( S\left( E\right)  \) has the property \( A \),
\( B \) or \( C \). In codimension 0 (no external parameter), a
generic curve \( S\left( E\right)  \) will be characterized by the
bounds of these intervals. In the next section, we will follow the
same scheme to systematically classify all these bounds, thus all
codimension 0 singularities.

\subsection{Codimension \protect\protect\( 0\protect \protect \) singularities\label{sec_codim0}}

In this section, we
construct all \emph{generic entropy curves}, that is curves accessible
without tuning an external parameter.
We call codimension \( 0 \) points the special points met along a
generic \( S(E) \)
curve when varying the energy and no other parameter. They are of
codimension \( 0 \) with respect to one of the three sources of singularities
(the maximization process, the convexity properties of \( S \) and
the convexification of \( S \)), and generic points with respect to
the other two: roughly speaking, we can ``use'' the tuning of the
energy to find a special point with respect to just one of the
sources. Thus, we first enumerate the codimension \( 0 \)
singularities arising from each of the three sources, and then combine
them with the generic points \( A \), \( B \) or \( C \).

\subsubsection{The three sources of singularities}

\begin{itemize}
\item \textbf{Maximization singularities and analytic properties}\\
  We want to classify all the singularities for the maximum of a
  variational problem with a constraint, when the value of the
  constraint is varied. We do not know any rigorous results dealing
  with this problem, in the literature. However, we use results from
singularity theory~\cite{Brysgalova}, proved for variational problems
depending on a parameter. Here we have no parameter in the function
but its role is played by the energy constraint. This difference might
be important in some particular cases; we neglect this possibility in
the following.  Thus, from~\cite{Brysgalova}, we know that codimension
\( 0 \) maximization singularities are of only one type. It
corresponds to an exchange of stability/metastability between two
different branches of solution of the variational problem. At such
point \( E_{c} \), \( S \) is continuous, but the derivative \( \beta
=\partial S/\partial E \) undergoes a \emph{positive} jump.  We will
refer to these points as \emph{microcanonical first order
  transitions}.
\item \textbf{Convexity properties}\\
 There is only one type of codimension \( 0 \) singularity arising
from the convexity properties of \( S \): the inflexion point, where
the second derivative of \( S \) vanishes (thus \( d\beta /dE=0 \)),
and the third derivative is non zero. Since this corresponds to the
point where a local minimum of \( f(m,\beta )=\beta h(m)-s(m) \)
becomes a saddle point, we will refer to it as a \emph{canonical
spinodal point} \footnote{We note that due to the long range interactions, the crossing of a spinodal point is not associated to a spinodal decomposition (phase separation and coarsening) as in short range interacting systems. The phenomenology of the dynamics will rather be associated to some global destabilization.}.
\item \textbf{Properties with respect to convexification}\\
 The concave envelop of \( S \) is made by concave portions joined
by straight lines. Thus, the only codimension \( 0 \) singularity
arising from convexification is the junction of a curved and a straight
portion of the concave envelop. Since this corresponds to a jump in
the first derivative of the free energy \( f(\beta ) \), we will
refer to it as a \emph{canonical first order transition}. At these
points, two canonical equilibrium states correspond to the same
$\beta$, and only one of those is the microcanonical state; this 
situation is also called partial equivalence of ensembles~\cite{ellis}. 
\end{itemize}

\subsubsection{Construction of the codimension \protect\protect\( 0\protect \protect \)
singularities}

We are now ready to construct all codimension \( 0 \) singularities
of the \( S(E) \) curve, by combining a codimension \( 0 \) situation
in one of the three sources of singularities with a generic point
in the other two. 

\begin{itemize}
\item Let us first consider the \emph{microcanonical first order transition}, see Fig.~\ref{codim0a}.
At the point where two branches of \( S\left( E\right)  \) meet,
the jump in the derivative \( \partial S/\partial E \) is necessarily
positive, so that the angle is reentrant. The point thus never belongs
to the concave envelop of \( S \): it is never visible in the canonical
ensemble, it is in the \emph{inequivalence range}. These two branches
have a definite concavity, and may both be concave or convex. This
determines three types of microcanonical phase transitions, the
concave-convex and convex-concave ones being equivalent by change
$E\to-E$.  
\item Let us now turn to the inflexion point, or \emph{canonical
    spinodal point}: it connects a concave and a convex portion of \( S \),
and cannot belong to the concave envelop of \( S \): it is in
the \emph{inequivalence range.} There is only one type of such points,
represented on Fig.~\ref{codim0a}. 
\item The \emph{canonical first order}
      transition bridges two contact point of the double
      tangent to the curve $S(E)$. At these two contact points, the \( S(E) \) curve is always locally concave and there is only one type of such
      points, see Fig.~\ref{codim0a}.
\end{itemize}
This completes the classification of codimension \( 0 \)
singularities.  To summarize, there exist microcanonical first order
transitions (three different types, according to the concavity of the
branches), canonical first order transitions (one type), and canonical
spinodal points (one type). From these, only the canonical first order
transition is visible in the canonical ensemble; the others belong to
the inequivalence range. A generic codimension 0 \(
S\left( E\right) \) curve is constructed with A, B or C type segments
separated by these singularities.

\subsubsection{Examples in physical systems}
Hertel and Thirring introduced in a pioneering paper an exactly
solvable model~\cite{hertel} to illustrate the concepts of negative
specific heat and inequivalence of ensembles. This simple model
already shows the three types of generic points (including the
negative specific heat), and all codimension 0 singularities: a
\emph{canonical first order transition}, a \emph{canonical
  spinodal point}, and a \emph{microcanonical first order transition}
between a convex and a concave branch. These situations are also found
in many different versions of more realistic
self-gravitating systems \cite{Chavanis_Houches,ispolatov,millerbruce,Stahl_Kiessling_Schindler}.\\
Examples can be found in other branches of physics:
in~\cite{Ellis_Stabilite}, Ellis and collaborators show a \( \beta
\left( E\right) \) curve for the quasi-geostrophic model displaying
the three types of generic points, a \emph{canonical first order
  transition}, and a \emph{canonical spinodal point};
in~\cite{Kiessling_Neukirch} Kiessling and Neukirch find the same
phenomenology in a magnetically self-confined plasma torus.\\
All codimension 0 situations have thus already been found in physical
models; we turn now to the the codimension~1 situations.

\subsection{Varying an external parameter: codimension 
\protect\protect\( 1\protect \protect \)
singularities \label{sec_codimension1}}

The codimension \( 1 \) singularities arise along the \( S(E) \) curve
when varying the energy and one external parameter: they describe the
way a generic $S(E)$ curve can be modified into another one. We
follow the same path as in the previous section to classify them:
using singularity theory, we first enumerate all codimension
\( 1 \) situations for any of the three sources of singularities,
independently of the two other sources. We then construct all 
codimension \( 1 \) singularities by combinations: either of a
codimension \( 1 \) situation in one source with a generic point in
the other two, or of a codimension \( 0 \) situation with respect to
two sources with a generic one in the last source.

\subsubsection{Codimension 1 singularities for the three sources of singularities}

\begin{itemize}
\item \textbf{Maximization singularities and analytic properties}\\
 Using results of singularity theory, we know that the codimension
\( 1 \) maximization singularities are of three types~{\cite{Brysgalova}}. 
\begin{itemize}
\item the first type is the crossing of two microcanonical first order transitions.
It happens when three different branches of solutions of the variational
problem have the same entropy. We will refer to it as a \emph{microcanonical
triple point}. 
\item the second type is the appearance of a microcanonical first order
transition; it corresponds to a point where \( d\beta /dE=d^{2}S/dE^{2}=+\infty  \).
We will call it a \emph{microcanonical critical point}. 
\item the last type consists in the simultaneous appearance of two microcanonical
first order transitions; it happens when a formerly wholly
microcanonically metastable
branch of \( S(E) \) crosses the stable one. Following the terminology
used in binary mixture phase transitions, we will call it an \emph{azeotropic
point} (see \cite{Aicardi2001}, and Fig.~\ref{codim1} for an illustration). 
\end{itemize}
\item \textbf{Convexity properties}

\begin{itemize}
\item Varying the energy and an additional parameter allows one to pick
up on the \( S(E) \) curve points where the second and third derivatives
of \( S(E) \) vanish. They show up as inflexion points with a horizontal
tangent in the \( \beta (E) \) curve. This situation will be called
a \emph{convexity change}. 
\end{itemize}

\item \textbf{Properties with respect to convexification}\\
 When varying an external parameter, the structure of the concave
envelop, as a succession of concave regions and straight lines, changes.
The codimension \( 1 \) singularities arising from convexification
are precisely the points where this structure changes; they are of
two types: 

\begin{itemize}
\item The first one is the appearance of a straight portion in the concave
envelope in a formerly concave region. This requires that
 \( d^{2}S/dE^{2} \)
and \( d^{3}S/dE^{3} \) vanish together. It is actually a convexity
change as described above, but with the additional request to be in
the concave envelop. This type of convexity change is the \emph{canonical 
critical point}. 
\item The second one is the breaking of a straight portion in two straight
portions separated by a concave zone. This happens when there is
a triple tangent to the \( S(E) \) graph; this will be called a 
\emph{canonical
triple point}. 
\end{itemize}
\end{itemize}

\subsubsection{Construction of the codimension \protect\protect\( 1\protect \protect \)
singularities}

The codimension \( 1 \) singularities are constructed by combining
a codimension \( 1 \) situation from one of the three sources of
singularity, with a generic point for the other two, or by combining
two codimension \( 0 \) singularities. 

\begin{itemize}
\item The \emph{microcanonical triple point} involves three different
  branches of solution, that may all be either convex or concave.  We
  refer them by three letters, for instance CVC (C for concave and V
  for convex). The first letter refers to the convexity of the low
  energy branch, the second one to the convexity of the appearing
  branch, and the third one to the convexity of high energy branch
  (see Fig.~\ref{codim1}). There are eight types of such points,
  however the CCV and VCC or the CVV and VVC cases, respectively, are
  equivalent (by a change $E \rightarrow -E$). The six remaining cases
  are illustrated on Fig.~\ref{codim1}. All these situations are
  always invisible in the concave envelop, thus
    always canonically invisible.
\item The \emph{microcanonical critical point} necessarily arises, in
  codimension 1, inside a convex region of the \( S(E) \) curve,
  invisible in the concave envelop. Thus, there is only one type of
  such points, see Fig.~\ref{codim1}.
\item The \emph{azeotropy} phenomenon involves two branches, but it is not
possible to have the lower branch convex and the upper one concave.
This leaves three cases, from which only the concave-concave (CC) one
is visible in the canonical ensemble; see Fig.~\ref{codim1}. We
refer to these situations by two letters, for instance CV, where the
second letter refers to the appearing branch. 
\item There are two types of \emph{convexity change}: one corresponds
  to the appearance
of a convex zone inside a concave one (referred by C), and the other
of a concave zone in a convex one (referred by V). The first case
may be visible after convexification. When such it is the \emph{canonical
critical point}. See Fig.~\ref{codim1}. 
\item The \emph{canonical triple point}, or triple tangent case, always
connects three concave parts of the \( S(E) \) curve, so that there
is only one type of such points, see Fig.~\ref{codim1}. 
\item We turn now to a combination of two codimension \( 0 \)
  singularities: the \emph{encounter of} a \emph{canonical
    spinodal point with a microcanonical first order transition}. Let
  us suppose that the branch with the inflexion point is the low
  energy one (other cases are recovered by the change $E \rightarrow
  -E$). The four cases CV-C, CV-V, VC-C and VC-V exist, where the two
  first letters refer to the concavity of the branch with the
  inflexion point, and the last one, to the concavity of the other
  branch, see Fig.~\ref{codim1}.\\
  The canonical first order transition always happens alone, so that
  there is no other possible combination of codimension \( 0 \)
  situations.
\end{itemize}
The codimension \( 1 \) singularities allow us to discuss the onset
of ensemble inequivalence, ie how the concavity of the \( S(E) \)
curve is destroyed when varying an external parameter. The inspection
of the above classification tells us that this can happen in only two
ways: at a canonical critical point and at an azeotropic point. 

Let us also summarize the links between microcanonical and canonical
codimension 1 singularities. The canonical critical point corresponds
to the appearance of a convex intruder in the entropy, the canonical
triple point is linked to a convexification singularity, whereas the
canonical azeotropy occurs together with a CC microcanonical azeotropy. 

We give now examples of physical models displaying some of these 
codimension $1$ situations.

\subsubsection{Examples in physical systems}

A cut-off is often imposed to regularize the short-range singularity
of self gravitating systems; the tuning of this new parameter allows
to observe codimension 1 singularities. The first rigorous study of the free energy of self gravitating fermions has been performed by Hertel and Thirring~\cite{Hertel71}, and the illustration of ensemble inequivalence associated to a first order canonical phase transition is presented in~\cite{Thirring_book}. A comprehensive description of the phenomenology of the phase diagram may be found in~\cite{Chavanis_Houches}: in addition to all types of codimension 0 singularities, the phase diagram exhibits a microcanonical critical point, a canonical critical point, and a
crossing point between a microcanonical first order transition and a
canonical spinodal point. In section \ref{sec_caracterisation},
we argued that the knowledge of the entropy-energy curve, in
particular of its concavity and convexification properties, is
sufficient to characterize both the microcanonical and canonical
phase transitions. In order to illustrate this point we draw a
schematic \( (-E,r ) \) phase diagram for self-gravitating
fermions ($r$ is the short range cut-off) 
on figure \ref{Fig_Diagramme_Phase_Autogravitant}. On this
diagram, both microcanonical and canonical phase transitions are
represented.

A similar phenomenology has been found in
the study of many different versions of self gravitating systems, the
external parameter being always the short distance cut-off imposed on
the gravitational interaction in various 
ways~\cite{Chavanis_Houches,Stahl_Kiessling_Schindler}. We refer
to the article \cite{Chavanis_PRE_2002} for a more complete
bibliography and a more comprehensive description of phase transitions
in self-gravitating systems.

To summarize, if all codimension 0 singularities have been found in
various models of interacting particles, just a few codimension~1
situations have already been observed. The classification thus allows
to point out new possible phenomenologies: \emph{microcanonical} and
\emph{canonical triple points}, \emph{azeotropy}, and \emph{convexity
  change V} (that is appearance of a 
concave zone inside a convex one) have
to our knowledge never been described in the literature on long range
interacting systems. Inspection of these transitions shows that not only
negative specific heat may occur in systems with long range interaction,
but also a wide zoology of uncommon behaviors.\\

\section{Classification of phase transitions and ensemble inequivalence situations
in systems with symmetry\label{sec_Classification_Symetrie} }

The classification of the previous section succeeded in reproducing
the phenomenology of many different systems, but not all of them.  In
particular, the various situations associated with second order phase
transitions do not appear. The reason
is that the notion of codimension \( 0 \), codimension \( 1 \)
singularities and so on were introduced under the only hypothesis that
the functional \( s(m) \) is infinitely differentiable, without any
reference to the physical system considered. However, the symmetries
of the physical situation, if present, will reflect in the functional
\( s(m) \), which then has to verify a priori some additional
hypothesis. This will lower the codimension of some singularities,
creating a much richer phenomenology already at the level of
codimension \( 0 \) and \( 1 \) singularities. Thus, we have to
complete our classification to apply it to physical systems with
symmetries, like the parity symmetry of an Ising spin
  system, the rotational symmetry of a self gravitating system in a
  spherical box...


\subsection{Relation between symmetry and codimension on a simple example}
\label{sym_codim_example}

To make clearer the point raised in the previous paragraph, let us
consider the mean field Ising model. Without external
  field, it displays a second order phase transition in the canonical
  ensemble, not described in the classification so far: this is
  directly related to the parity symmetry of the model. With an
  external field, this symmetry is destroyed, as well as the second
  order phase transition; a first order phase transition, codimension
  0 singularity even in the absence of symmetry, may however remain.

To be more precise, we consider a functional \( s(m,E) \)
with \( m\in {\mathcal{R}} \), with a parity symmetry: \( s(-m,E)=s(m,E) \).
As a consequence, all derivatives of odd order vanish in \( m=0 \): 
\( \partial ^{(2n+1)}_{m}s|_{(0,E)}=0 \).
Let us consider now a microcanonical
critical point, defined as a point \( (m_{c},E_{c}) \) where \( \partial ^{(n)}_{m}s|_{(m_{c},E_{c})}=0 \),
for \( n=1,2,3 \). Without any hypothesis on the function \( s \),
this requires a priori to satisfy three equations. Thus, a generic
critical point can only be found by adding another degree of freedom, 
besides $m$ and $E$,
and is of codimension \( 1 \). For instance a normal form for a critical
point is given by \( s\left( E,\lambda \right) =\max _{m}\left\{ -m^{4}+bm^{2}+am\right\}  \)
where \( a \) and \( b \) are linear combination of \( E \) and
\( \lambda  \),  and $\lambda$ is a tunable external parameter.
The critical point is given by \( a=b=0 \) and the
line of microcanonical first order transition is given by $a=0$ 
and $b>0$. However, if \( s \) is symmetric under parity, all odd derivatives
automatically vanish at \( m=0 \), so that finding a critical point
breaking this symmetry requires only to satisfy one equation, varying
\( E \). Moreover as the first derivative identically vanishes, the
transition is continuous: a microcanonical second order phase transition
is in that case a codimension \( 0 \) phenomenon. For instance, a
normal form is given by \( s\left( E\right) =\max _{m}\left\{ -m^{4}+\left( E-E_{C}\right) m^{2}\right\}  \).

In the following, we make the hypothesis that for a more general symmetry,
at least one variable \( m_{2} \) may be found such that all odd 
derivatives with respect to \( m_{2} \) of \( s(m_{1},m_{2},E,\lambda ) \)
identically vanish: \( \partial ^{(2n+1)}_{m_{2}}s|_{m_{2}=0}=0 \). This is 
actually the case for a rotational symmetry, where the radius \( r \) plays
the role of \( m_{2} \). A case by case study should
be done for any other symmetry. 

We are not aware of any mathematical result we could use to enumerate
the new singularities, such as \cite{Brysgalova} for the non symmetric
case. We will thus use in the following the heuristic, but
systematic, criterion based on the ``number of equations to be
solved'', as explained in this paragraph.

\subsection{Codimension \protect\protect\( 0\protect \protect \) singularities}


The singularities associated to the convexity properties, as well as
those concerning the convexification, are not modified when a symmetry
is assumed. As seen above however, there is now an additional
singularity due to the maximization process: the second order phase
transition, associated with a symmetry breaking. At these points, a
microcanonically stable branch obeying the symmetry
loses stability, and a non symmetric stable branch appears. By
convention, we will suppose that the symmetric branch is stable at
energies larger than \( E_{c} \), and the broken symmetry branch is
stable for energy smaller than \( E_{c} \), although it is not
necessary. \( S(E) \) and \( dS/dE \) are continuous at \( E=E_{c} \),
but \( d^{2}S/dE^{2} \) experiences a \emph{negative jump} (a positive
jump if the opposite convention were adopted).

A priori, the high and low energy branches may be concave or convex,
but because of this negative jump condition, the association of a
concave low energy branch with a convex high energy branch is
impossible. We are thus left with three types of second order phase
transitions denoted CC, VC and VV where the first letter refers to the
low energy branch convexity, see Fig.~\ref{codim0_Symmetrie}. Only the
CC microcanonical second order phase transition is visible in the
canonical ensemble: it is the canonical second order phase transition.
This completes the classification of codimension \( 0 \)
singularities.


\subsection{Codimension \protect\protect\( 1\protect \protect \) singularities}

The new codimension \( 1 \) singularities arise on one hand
from the new codimension \( 1 \) singularities due to the maximization,
and on the other hand from the combination of the second order phase
transition (of codimension \( 0 \)) with other codimension \( 0 \)
singularities. We first identify the new codimension \( 1 \) singularities,
due to symmetry.

\subsubsection{Codimension \protect\protect\( 1\protect \protect \) singularities
for the three sources of singularities}

\begin{itemize}
\item \textbf{Maximization singularities and analytic properties}\\

 The microcanonical critical point, microcanonical triple point and
azeotropic point are unchanged. Five new singularities are added. 

\begin{itemize}
\item the first one appears when the first five derivatives of \( s \)
  vanish together along a certain direction. A normal form is given by
  \( S\left( E,\lambda \right) =\max _{m}\left\{
    -m^{6}-3bm^{4}/2-3am^{2}\right\} \), where \( a \) and \( b \) are
  linear combinations of \( E \) and \( \lambda \), and $\lambda$ is
  the tunable external parameter which gives access to codimension~1
  singularities. In the \( (E,\lambda ) \) plane, it connects a line
  of microcanonical first order transitions with a line of second
  order phase transitions; following the usual canonical terminology
  we will refer to it as a \textbf{microcanonical tricritical
    point}~\cite{Domb}. Fig.~\ref{Fig_tricritique} shows the
  transition lines in the $(E,\lambda)$ plane in the vicinity of a
  microcanonical tricritical point. The qualitative features of this
  diagram are universal: they do not depend on the polynomial form
  chosen for~$S$.
  
\item the second one occurs when \( s \) has two equal maxima, and one
  of these is quartic along one direction. A normal form is given by
  \( S_{1}\left( E,\lambda \right) =\max _{m_{1}}\left\{ -m_{1}^{4}-
    2am_{1}^{2}\, ;\, -m_{1}^{2}+b\right\} \), where \( a \) and \( b
  \) are linear combinations of \( E \) and \( \lambda \). 
  In the \( (E,\lambda ) \)
  plane, it connects a line of microcanonical first order transitions
  with a line of second order phase transitions. It is a
  \textbf{microcanonical critical end point}~\cite{Domb}. Transition
  lines in the vicinity of a microcanonical critical end point are
  shown on Fig.~\ref{Fig_crossing_order1_order2}.

  
\item the third and fourth happen at a crossing of two second order
  transition lines, when the maximum of \( s \) is quartic along two
  directions having the symmetry property. A normal form is given by
  \( S\left( E,\lambda \right) =\max _{m_{1},m_{2}}\left\{
    -m_{1}^{4}-2am_{1}^{2}-m_{2}^{4}-2bm_{2}^{2}-cm^{2}_{1}m^{2}_{2}\right\}
  \), where \( a\rightarrow 0 \) and \( b\rightarrow 0 \) are linear
  combinations of \( E \) and \( \lambda \), and \( c \) is a finite
  constant
  with \( c>-1 \). Two cases have to be considered (see Fig.~\ref{Fig_crossing_order2_order2}).\\
  When \( c<1 \), varying the external parameter \( \lambda \), two
  second order transitions come closer and closer; once the critical
  value of the parameter is reached, they cross each other and remain
  unaffected. We are not aware of any standard denomination for this
  singularity; as it involves four phases (\( m_{1}=m_{2}=0 \), \(
  m_{1}=0\, \, \textrm{and}\, \, m_{2}\neq 0 \), \( m_{1}\neq 0\, \,
  \textrm{and}\, \, m_{2}=0 \), and \( m_{1}\neq 0\, \, \textrm{and}\,
  \, m_{2}\neq 0 \)), we call it a \textbf{second order quadruple
    point}.  When \( c>1 \), the doubly asymmetric phase \( m_{1}\neq
  0\, \, \textrm{and}\, \, m_{2}\neq 0 \) is unstable (see
  Fig.~\ref{Fig_crossing_order2_order2}). For \( a<0 \) and \( b<0 \),
  the two phases \( m_{1}=0\, \, \textrm{and}\, \, m_{2}\neq 0 \), \(
  m_{1}\neq 0\, \, \textrm{and}\, \, m_{2}=0 \) may have the same
  entropy, and a first order transition between them occurs when \(
  a=b \). It is a \textbf{microcanonical bicritical
    point}~\cite{Domb}.  Transition lines for all these situations are
  shown on Fig.~\ref{Fig_crossing_order2_order2}.

\item the last new singularity is the simultaneous appearance of two second
order phase transitions. It happens when the coefficient of the quartic
term around the maximum of \( s \), along a certain direction, has
a double root for \( e \). We will refer to it as a \textbf{second
order azeotropic point}. If we call \( \lambda  \) the free parameter,
the locus of a second order phase transition is represented by a curve
in the \( (\lambda ,E) \) plane; a second order azeotropy occurs
when a line of constant \( \lambda  \) is tangent to this curve.
A normal form is given by \( S\left( E,\lambda \right) =\max _{m}\left\{ -m^{4}+2\left( E^{2}-\lambda \right) m^{2}\right\}  \)
(see Fig.~\ref{Fig_second_order_azeotropy}).
\end{itemize}
\item \textbf{Convexity properties}\\
 No new singularity arises from the inspection of the convexity properties. 
\item \textbf{Properties with respect to convexification}\\
 No new singularity arises from the convexification of \( S(E) \). 
\end{itemize}

\subsubsection{Construction of the codimension \protect\protect\protect\( 1\protect \protect \protect \) singularities}
\label{sec_codim1_sym}

We now have to construct the new codimension \( 1 \) singularities by
combining a new codimension \( 1 \) situation for the maximization
process with a generic point for the other two sources of
singularities, or by combining a second order phase transition with
another codimension \( 0 \) situation. Although the procedure is
straightforward, discussing all cases is tedious. We have chosen to
give a detailed discussion for the \textbf{tricritical} and \textbf{bicritical
  points} only, in both ensembles. Table~\ref{Fig_codim1_sym}
summarizes all the results.

\paragraph{Tricritical points:} 

Inspection of the entropy development around $m=0$ for the
\textbf{microcanonical tricritical point} shows that at the
tricritical point $\lambda=\lambda_c$, $E \to E_c^-$, the temperature is
given by $\beta(E)\simeq -C(E_c-E)^{1/2}$, where $C$ is a positive
constant. This proves that the low energy branch (less symmetric phase
with our convention) is necessarily convex. According to the concavity
of the high energy branch, there are thus two types of such points,
which we denote VV and VC. Because of the convexity of at least one of
the branch, both types are invisible in the canonical ensemble. See 
Fig.~\ref{Fig_Tricritique2}.\\
We can construct a codimension~1 situation by superposing a second
order transition (codimension~0) with a canonical spinodal point in
one of the two
branches (codimension~0). If the inflexion point is in the
low energy branch, the second order transition evolves from a CC situation
to a VC one. This is visible in the canonical ensemble. As it corresponds
to the appearance of a first order transition from a second order
one, it is a \textbf{canonical tricritical point}. As a first order
canonical transition appears at this point, this singularity is associated
to the onset of ensemble inequivalence, see Fig.~\ref{Fig_Tricritique2}.
If the inflexion point is in the high energy branch, the situation
evolves from VV to VC; it is not visible in the canonical ensemble. 

\paragraph{Bicritical points:}
We base our discussion of the \textbf{microcanonical bicritical point}
on Fig.~\ref{Fig_crossing_order2_order2}d. Depending on the
orientation of the $E$ and $\lambda$ axis in the $(a,b)$ plane, there
are two ways to cross the singularity.  Along the first one ($E_1$,
solid line on Fig.~\ref{Fig_crossing_order2_order2}d), the first order
phase transition disappears at the singularity point, and the
succession of phases is Aa-Ab-S$\to$Aa-S ; we still adopt the
convention that the symmetric phase is the one of highest energy).
Taking into account the two negative concavity jumps associated with
the second order transitions, as well as the direction of $E$ in the
$(a,b)$ plane , we are left with four possible concavity
configurations for this Aa-Ab-S\( \rightarrow \)Aa-S singularity:
VVV\( \rightarrow \)VV, VVC\( \rightarrow \)VC, VCC\( \rightarrow
\)VC, and CCC\( \rightarrow \)CC (the CVC$\to$CC case is eliminated by
the choice of the direction for $E$: indeed, $S_{Aa}=a^2$,
$S_{Ab}=b^2$, and the choice of the $E$ direction is such that Ab is
more concave than Aa).  Only the last one (CCC\( \rightarrow \)CC) is
visible in the canonical ensemble. As it involves a
canonically invisible range near the first order
transition, it is associated with the ensemble inequivalence onset.
In the canonical ensemble, this situation involves a first order and a
second order phase transitions before the singularity and a second
order one after it. From the explicit computation of $F\left( \beta
\right)$ from the normal form, we conclude that the second order
transition before the transition is visible in the
canonical ensemble.  This singularity is thus a \textbf{canonical
  bicritical point}; see Fig.~\ref{Fig_Bicritique}
for schematic entropic and caloric curves in that case.\\
Along the $E_2$ direction (dashed line on
Fig.~\ref{Fig_crossing_order2_order2}d), the first order phase
transition disappears and two second order phase transitions take
place: Aa-Ab\( \rightarrow \)Aa-S-Ab. Taking into account the two
concavity jumps associated with the second order transitions, we have
four possible concavity configurations for this Aa-Ab\( \rightarrow
\)Aa-S-Ab type: VV$\to$VVV, VV$\to$VCV, VC$\to$VCC, CC$\to$CCC (the
case CV$\to$CCV is also possible, but equivalent, by symmetry, to the
VC$\to$VCC one).  Only the CC$\to$CCC case is canonically visible, and
it is associated with ensemble inequivalence onset. It is also a {\bf
  canonical bicritical
  point}. \\

 The same discussion is carried out for all other singularities in appendix.
This allows to enumerate all possible microcanonical and canonical
transitions and their mutual relationship. The results are summarized
in table~\ref{Fig_codim1_sym}. Most of the new singularities take
place in the inequivalence range, and thus are canonically invisible,
but two new types of inequivalence onset are found, associated with
the canonical tricritical and bicritical points.\\
Let us comment further on the links between canonical and
microcanonical ensembles. As the canonical optimization problem and
the microcanonical one are formally equivalent (once the energy
constraint is considered as a parameter in the microcanonical problem,
which is what we have done), we should find exactly the same
singularities in the two ensembles, even though we have constructed
the canonical solution by convexification of the microcanonical one
and not directly from the optimization problem. This is indeed what we
find, proving the consistency of our results.  All microcanonical
singularities have thus a canonical equivalent, but the microcanonical
phenomenology is much richer, as it allows many more concavity
configurations. We review in the following paragraph the different
situations already observed in the literature, and point out the
ones which have not been found yet.

\subsection{Examples}
 \label{sec_exemples_sym}
 One of our main goals is to study the onset of ensemble
 inequivalence, or equivalently
 the failure of the concavity of \( S(E) \) varying an external parameter.
 For non symmetric systems, the canonical critical point and the
 azeotropic point have been shown to be the only two types of
 inequivalence onset. Two new types are identified for systems with
 symmetry. One is the canonical tricritical point, across which a CC
 microcanonical second order transition becomes a VC one.  Onset of
 inequivalence through a {\bf canonical tricritical point} has been
 observed at least twice, in a toy model of self gravitating
 system~\cite{Antoni2} and in a spin system~\cite{barre}. This
 situation has also been studied in a general setting~\cite{Leyvraz}.
 Onset of inequivalence around a \textbf{canonical bicritical point}
 has to our knowledge never been found.\\
 Concerning the other codimension \( 1 \) singularities listed in this
 section, we are aware of just a few references. A
 \textbf{microcanonical tricritical point} with a concave high energy
 branch has been found in a spin system~\cite{barre}; a
 \textbf{microcanonical critical end point} has been observed in
 \cite{Antoni2}. All the other classified situations, some of which
 are rather exotic, have apparently never been found. As for the non
 symmetric case, we expect some of these new situations to be
 found when more complex models are studied. Let us note that some of
 these singularities have of course been found in a purely canonical
 setting: see for instance~\cite{Krinsky75a}.\\

\section{Conclusion\label{sec_Conclusion}}

We have proposed in this article a classification of phase transitions
in non additive systems. It relies on the fact that these systems
are ``mean field like'', which enables one to find the microcanonical
and canonical equilibrium states as solutions of variational problems. 
Taking advantage of this structure, and making use of existing results from
singularity theory when available, or of heuristic arguments when not,
we have classified all phase transitions for these systems up to codimension
1, that is we have enumerated all the elementary pieces building the
phase diagrams of systems with one constraint (usually the energy)
and one free parameter.
All these results are summarized by Figs.~\ref{codim0a}, \ref{codim1}, 
\ref{codim0_Symmetrie} and table \ref{Fig_codim1_sym}.\\
This classification gives a unique framework to understand the unusual
thermodynamic phenomena arising in the many fields of physics concerned
by this non additivity problem: astrophysics, two dimensional and
geophysical turbulence, some models of plasma physics. All phase diagrams
obtained so far in these fields (analytically or numerically), are
reproduced by the classification. In addition, we have exhibited some
phenomenologies likely to appear in the non additive context, but
not yet found in any specific model. Among them, we can mention
azeotropy and canonical bicritical points, which are new types of ensemble
inequivalence onset. Moreover, lots of singularities associated with precise
convexity and convexification properties have also not been observed.
We have not emphasized in the paper the critical exponents of the various
transitions; however, it is clear that since the mean field approach is 
valid, they are totally universal for the systems considered, and may be 
calculated easily.\\
We have restricted in this article the classification to codimension
1 situations (corresponding to systems with one free parameter), and
one conserved quantity (the energy in the whole article). It is thus
possible to generalize this work, either by classifying situations
of codimension greater than 2, or, maybe more interestingly, by considering
systems with two or more conserved quantities (angular momentum, total
circulation for fluids models). Taking into account more conserved
quantities is likely to give a much richer phenomenology, as shown
for instance in the work~\cite{Votyakov}, for the self gravitating
gas with short range cut off, and conserved angular momentum, in addition to 
the energy. Another problem
not analyzed in this work is the possibility of singularities at the border
of the accessible energy range for the system.\\

\section*{Acknowledgments}

We are very pleased to thank Pierre Henri Chavanis, Thierry Dauxois
and Stefano Ruffo for very useful discussions
and advice. We thank especially Pierre Henri Chavanis for 
some specific comments on the self-gravitating system example. 
J.~B. also thanks Matthew Hastings.\\
This work has been supported by the French Minist{\`e}re de la
Recherche through a Lavoisier fellowship, and by R{\'e}gion
Rh{\^o}ne-Alpes for the
fellowship N$^\circ$ 01-009261-01.\\

\begin{appendix}
\section*{Appendix}
In this appendix we give a detailed discussion of the construction of all codimension 1 singularities in a system with symmetry (see section \ref{sec_codim1_sym}), corresponding to the microcanonical critical end point, to the second order quadruple point, and to the second order azeotropic point.
  
\begin{itemize}
\item {\bf The microcanonical critical end point} is
    schematically represented on
    Fig.~\ref{Fig_crossing_order1_order2}. With axis
    $(E,\lambda)$ as drawn, there is one first order transition for
    $\lambda<0$, and a first order then a second order transitions for
    $\lambda>0$. Combining the three types of second order transitions
    with the two possible concavities of the additional branch yields
    six different situations, denoted C-CC, V-CC, C-VC, V-VC, C-VV and
    V-VV, where the first letter refers to the first order transition
    concavity and the two last ones refer to the second order
    transition. Other directions for $E$ and
    $\lambda$ on Fig.~\ref{Fig_crossing_order1_order2} are recovered
    by symmetry.


\item the \textbf{second order quadruple point} involves four phases,
  visible on figure \ref{Fig_crossing_order2_order2}, denoted AA, Aa,
  Ab, and S (A refers to asymmetric and S to symmetric, see figure
  \ref{Fig_crossing_order2_order2}a, b and c). Let us analyze the
  convexity of the entropy $d^2S(E)/dE^2$, where the energy $E$ is
  along a fixed direction in the $(a,b)$ plane. Each of the four
  second order transitions AA-Aa, AA-Ab, Aa-S and Ab-S involves a
  negative concavity jump, independently of the way the transitions
  are crossed in the $(a,b)$ plane. Taking into account the signs of
  these jumps, the 6 following cases are possible, for the concavity
  of the branches AA, Aa, Ab and S respectively: VVVV, VVVC, VVCC,
  VCVC, VCCC, CCCC. A priori, we do not have any information about
  which branch between Aa or Ab is more convex than the other;
  however, fixing
  the direction of $E$ in the $(a,b)$ plane may force the concavity
  jump between Aa and Ab to have a definite sign, as we will see.\\
  When crossing the singularity, the type of transition (succession of
  phase on the $E$ line) depends on the $E$ direction (see for
  instance $E_1$, $E_2$ and $E_3$ on figure
  \ref{Fig_crossing_order2_order2}). Along the $E_1$ direction, one
  goes from a succession of phases AA-Ab-S to AA-Aa-S; along the $E_2$
  direction, the succession is Ab-S-Aa \( \rightarrow \) Ab-AA-Aa;
  along $E_3$, the succession is AA-Ab \( \rightarrow \) AA-Aa-S-Ab
  (this is possible only if the parameter $c$ in the normal form is
  such that $-1<c<0$): three transitions appear from a single one;
  along $E_4$, the succession is Ab-S \( \rightarrow \) Ab-AA-Aa-S
  (this is possible only if the parameter $c$ in the normal form is
  such that $0<c<1$): also in this case, three transitions appear from
  a single one. All other possible directions for $E$ and $\lambda$
  can be recast into one of the four previous situations, using the
  changes of variables $E \to -E$, $\lambda \to -\lambda$, and the
  symmetric role played by the branches Aa and Ab. For instance, the
  $E_5$ direction on Fig.~\ref{Fig_crossing_order2_order2}c leads to
  Aa-AA-Ab $\to$ Aa-S-Ab, which is actually the same as $E_2$, once
  the changes $E\to -E$,
  $\lambda \to -\lambda$ are made.\\
  Taking into account the previously studied concavity configurations,
  5 possible AA-Ab-S \( \rightarrow \) AA-Aa-S transitions are found:
  VVV$\to$VVV, VVC$\to$VVC, VCC$\to$VVC, VCC$\to$VCC, CCC$\to$CCC (the
  VVC$\to$VCC case is also possible, but equivalent, by symmetry, to
  VCC$\to$VVC). Only the CCC$\to$CCC type transitions are visible in
  the canonical ensemble; it is not associated with the onset of
  ensemble inequivalence. Moreover, the continuity of \( \beta \left(
    E\right) \) and of \( S\left( E,\lambda \right) \) insures the
  continuity of \( F\left( \beta ,\lambda \right) \) and of \(
  \partial _{\beta }F\left( \beta ,\lambda \right) =-E\left( \beta
  \right) \).  Therefore the microcanonical AA-Ab-S \( \rightarrow \)
  AA-Aa-S, CCC\( \rightarrow \)CCC second order quadruple point is
  associated to a AA-Ab-S$\to$AA-Aa-S
  \textbf{canonical second order quadruple point.} \\
  Taking into account the concavity configurations, 5 possible
  Ab-AA-Aa \( \rightarrow \) Ab-S-Aa transitions are found:
  VVV$\to$VVV, VVV$\to$VCV, CVV$\to$CCV, CVC$\to$CCC, CCC$\to$CCC (the
  VVC$\to$VCC case is also possible, but equivalent, by symmetry, to
  CVV$\to$CCV). Only the CCC$\to$CCC and the CVC$\to$CCC types are
  canonically visible, and the latter is associated with the onset
  (disappearance when crossed that way) of an energy range where
  ensembles are not equivalent: the convex AA branch is replaced by a
  first order transition in the canonical one. The same reasoning as
  above for the continuity of \( F\left( \beta ,\lambda \right) \) and
  of \( \partial _{\beta }F\left( \beta ,\lambda \right) =-E\left(
    \beta \right) \) insures that the CC microcanonical second order
  transitions remain second order transitions in the canonical
  ensemble.  We conclude that the microcanonical Ab-AA-Aa \(
  \rightarrow \) Ab-S-Aa, CCC\( \rightarrow \)CCC second order
  quadruple point is associated to a Ab-AA-Aa$\to$Ab-S-Aa
  \textbf{canonical second order quadruple point}, and the
  microcanonical Ab-AA-Aa \( \rightarrow \) Ab-S-Aa, CVC\( \rightarrow
  \)CCC second order quadruple point is associated to a
  Ab-Aa$\to$Ab-S-Aa \textbf{canonical bicritical point}.\\
  Taking into account the concavity configurations, 5 possible AA-Ab\(
  \rightarrow \) AA-Aa-S-Ab singularities are found: VV$\to$VVVV,
  VV$\to$VVCV, VC$\to$ VVCC, VC$\to$VCCC, CC$\to$CCCC. The direction of
  $E$ in the vicinity of the singularity (for instance $E_3$ on
  Fig.~\ref{Fig_crossing_order2_order2}a) implies that the Aa branch
  is more convex than the Ab one; that's why the VV$\to$VCCV case has
  to be eliminated.  Only the CC$\to$CCCC type is canonically visible;
  it is not associated with ensemble inequivalence onset. We conclude
  that the microcanonical AA-Ab\( \rightarrow \) AA-Aa-S-Ab,
  CC$\to$CCCC second order quadruple point is associated with a
  AA-Ab\( \rightarrow \) AA-Aa-S-Ab {\bf canonical
    second order quadruple point}.\\
  Finally, taking into account the concavity configurations, 5
  possible Ab-S \( \rightarrow \) Ab-AA-Aa-S singularities are found:
  VV$\to$VVV, VC$\to$ VVVC, VC$\to$VVCC, CC$\to$CVCC, CC$\to$CCCC. The
  direction of $E$ implies that the Ab branch is more convex than the
  Aa one, so that we have eliminated the CC$\to$CVVC case. The
  CC$\to$CCCC case is canonically visible, not associated with
  ensemble inequivalence onset, and corresponds to a Ab-S \(
  \rightarrow \) Ab-AA-Aa-S {\bf canonical second order quadruple
    point}. The CC$\to$CVCC case is also canonically visible, and is
  associated to ensemble inequivalence onset. Before the singularity,
  the system shows canonically a second order phase transition, and,
  after the singularity, a first and a second order one. Thus, it is a
  Ab-S$\to$Ab-Aa-S \textbf{canonical bicritical point} (this is the
  canonical equivalent-upon symmetry- of the microcanonical
  Aa-Ab-S$\to$Aa-S bicritical point).

\item a \textbf{second order azeotropy} involves the appearance of two
  second order transitions. If we use the convention that the low
  energy state is an asymmetric one (as on figure
  \ref{Fig_second_order_azeotropy}), then just after the crossing of
  the singularity, one observes, varying the energy, two second order
  transitions, with the configuration asymmetric-symmetric-asymmetric.
  The jumps in \( d^{2}S/dE^{2} \) are thus exactly opposite. Moreover
  the jump is exactly zero at the transition point (the entropy change
  is there quartic). Thus the concavity is not changed at the
  transition point. This yields two types of such points, depending on
  the concavity of \( S \), denoted as VV and
  CC, with the usual convention. \\
  Only the CC case is visible in the canonical ensemble. In the
  canonical ensemble, the second derivative of the free energy with
  respect to \( \beta \) is singular, and we observe the appearance of
  two jumps at the singularity point. It is thus a \textbf{canonical
    second order azeotropy}.

\item a \textbf{second order transition may superpose with an inflexion
point in one of the two branches}. If the inflexion point is in the
low energy branch, the second order transition evolves from a CC situation
to a VC one. This is visible in the canonical ensemble. As it corresponds
to the appearance of a first order transition from a second order
one's, it is a \textbf{canonical tricritical point.} As a first order
canonical transition appears at this point, this singularity is associated
to the birth of an interval of energy for which microcanonical and
canonical ensembles are inequivalent.\\
If the inflexion point is in the high energy branch, the situation
evolves from VV to VC; it is not visible in the canonical ensemble. 
\item finally, a \textbf{concave-concave second order transition} \textbf{may
coincide with the boundary of a straight segment of the concave envelop}
of \( S(E) \). There is only one type of such point. It is canonically
visible. In the canonical ensemble, it gives a \textbf{}crossing between
a first order and a second order canonical phase transitions. It is
thus a \textbf{canonical critical endpoint.} It
appears at the boundary of an ensemble inequivalence range, but it
is not associated with ensemble inequivalence appearance.
\end{itemize}

\end{appendix}

\pagebreak
\section*{Captions for figures}
\begin{figure}[p]
\resizebox*{1\textwidth}{!}{\includegraphics{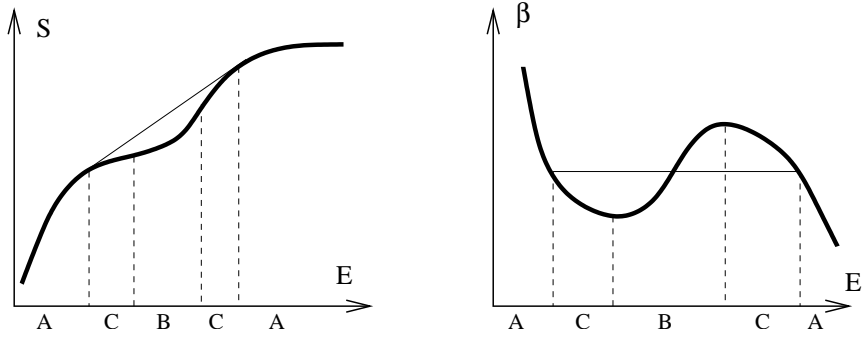}} 
\caption{ \label{codim-1}  The three types of generic points on the
  \protect\( S(E)\protect \) curve.  Points in A zones are concave and
  belong to the concave envelop: they correspond to a canonical
  solution; B points are convex, thus canonically unstable; C points
  are concave but do not belong to the concave envelop: they are
  canonically metastable. The thin line is the canonical curve in the
  inequivalence range.}  \vskip 10cm
\end{figure}

\noindent
Fig.~\ref{codim-1}: The three types of generic points on the
\protect\( S(E)\protect \) curve.  Points in A zones are concave and
belong to the concave envelop: they correspond to a canonical
solution; B points are convex, thus canonically unstable; C points are
concave but do not belong to the concave envelop: they are canonically
metastable. The thin line is the canonical curve in the inequivalence
range.\\

\begin{figure}[p]

\resizebox*{\textwidth}{!}{\includegraphics{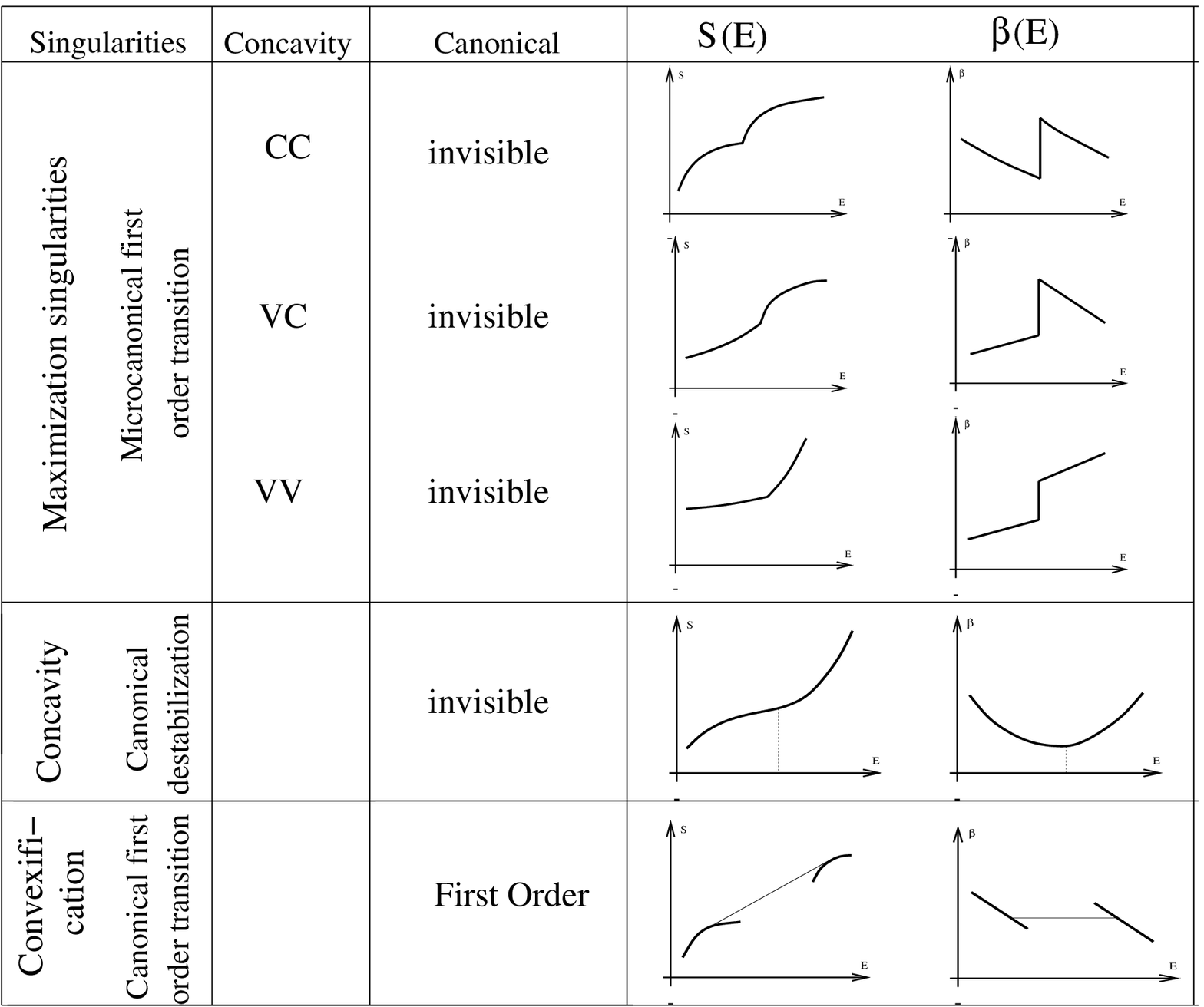}} 
\caption{\label{codim0a} Codimension 0 
  singularities. The first three rows are the three types of 
  microcanonical first order transition, the fourth is the canonical
  spinodal point, and the fifth is the canonical first order
  transition. C stands for Concave, and V for conVex. The bold line is
  the microcanonical solution as well as the canonical one when they
  are equivalent; the thin one is the canonical solution.
  ``Invisible'' means that the transition is not
  visible in the canonical ensemble. The canonical curve in not drawn
  for invisible situations.} 

\end{figure}

\noindent
Fig.~\ref{codim0a}: Codimension 0 
  singularities. The first three rows are the three types of 
  microcanonical first order transition, the fourth is the canonical
  spinodal point, and the fifth is the canonical first order
  transition. C stands for Concave, and V for conVex. The bold line is
  the microcanonical solution as well as the canonical one when they
  are equivalent; the thin one is the canonical solution.
``Invisible'' means that the transition is not
visible in the canonical ensemble. The canonical
curve in not drawn for invisible situations.\\

\begin{figure}[p]

{\centering \rotatebox{90}{\resizebox*{.9 \textheight}{1.3\textwidth}
{\includegraphics{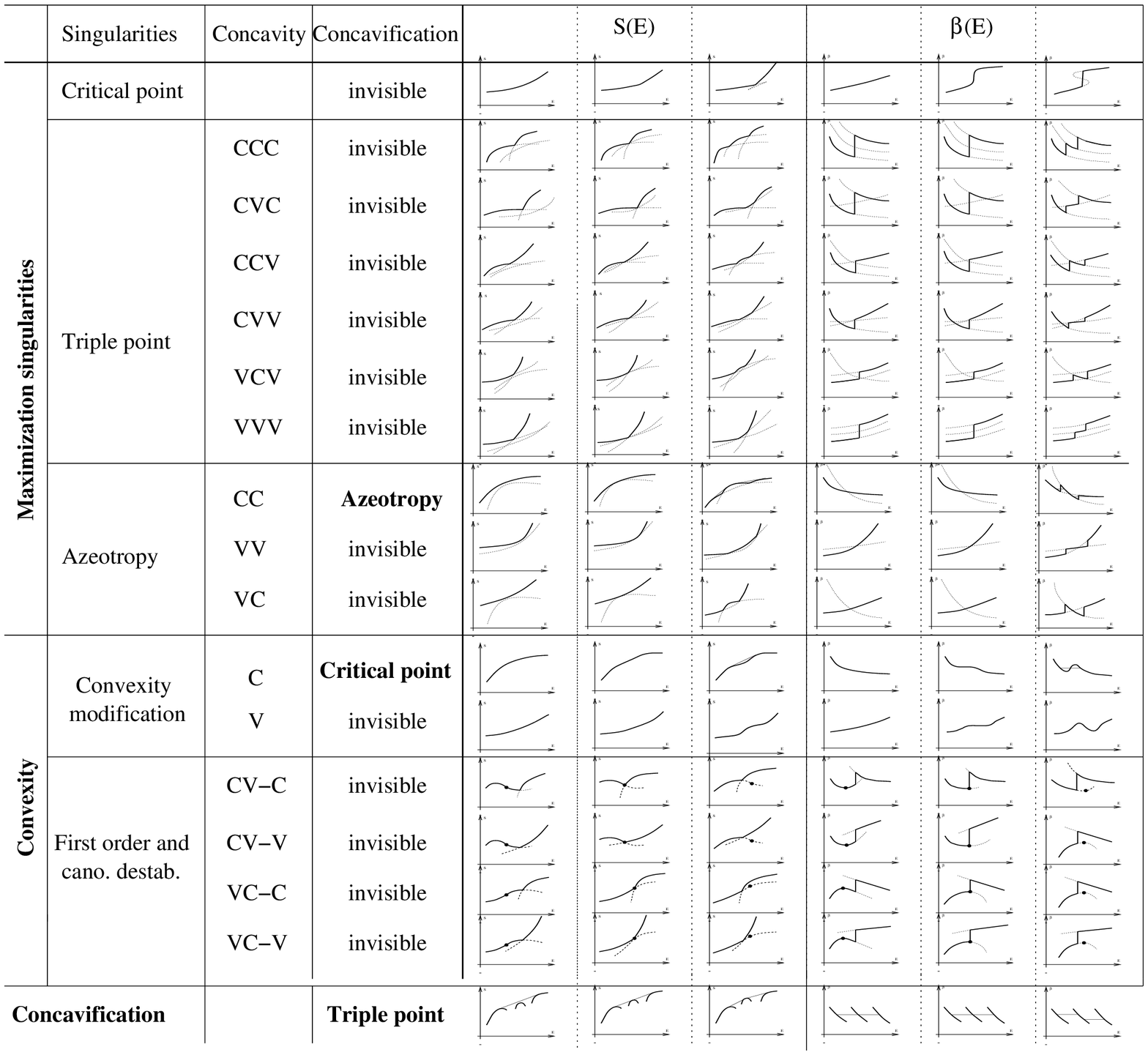}}} \par}
\caption{\label{codim1} Codimension 1 singularities. The three
    curves $S(E)$ and $\beta(E)$ for each singularity correspond to
    the situation just before the singularity is crossed, right at the
    singularity, and just after it.  See the text for comments; the
  meaning of the curves is as in Fig.~\ref{codim0a}, and dotted lines
  represent metastable or unstable microcanonical branches.
  }
\end{figure}

\noindent
Fig.~\ref{codim1}: Codimension 1 singularities. The three
  curves $S(E)$ and $\beta(E)$ for each singularity correspond to the
  situation just before the singularity is crossed, right at the
  singularity, and just after it.  See the text for comments; the
meaning of the curves is as in Fig.~\ref{codim0a}, and dotted lines
represent metastable or unstable microcanonical branches.\\

\begin{figure}[p]

{\centering \resizebox*{0.8\textwidth}{0.4\textheight}
{\includegraphics{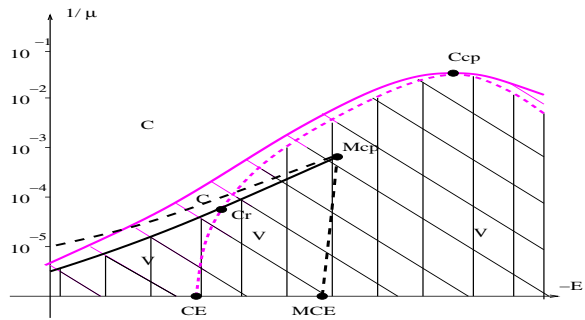}} \par}
\caption{\label{Fig_Diagramme_Phase_Autogravitant} Schematic phase diagram
for self-gravitating fermions. Horizontal and
  vertical axis are respectively minus the energy and a small length
  scale cut-off. This
illustrates that a microcanonical diagram including
microcanonical phase transitions as well as entropy concavity and
convexification properties is sufficient to summarize the canonical
phase transitions. In bold black, a line of first order microcanonical
phase transition (between gaseous and core-halo phases), ending
at a \textbf{microcanonical critical point (Mcp)}. In bold grey, a line
of canonical first order transition (convexification singularity),
with a \textbf{canonical critical point (Ccp)}. In dashed grey,
a line of canonical spinodal point. It crosses the
microcanonical first order phase transition at~Cr. The hashed zone is
the inequivalence area, absent in the canonical ensemble; the
doubly hashed is the negative specific region. The bold dashed line is a line of microcanonical stability change,
marking the limits of stability of the gaseous and core-halo phases
(such microcanonical stability changes are not described in our 
classification).  For \protect\( 1/\mu =0\protect \),
we recover the self-gravitating isothermal collapse (CE) and gravitational
phase transition (MCE) points. A similar phase
diagram has been done independently in \cite{Chavanis_Rieutord}.}
\vskip 3cm
\end{figure}

\noindent
Fig.~\ref{Fig_Diagramme_Phase_Autogravitant}: Schematic phase diagram
for self-gravitating fermions. Horizontal and
  vertical axis are respectively minus the energy and a small length
  scale cut-off. This
illustrates that a microcanonical diagram including
microcanonical phase transitions as well as entropy concavity and
convexification properties is sufficient to summarize the canonical
phase transitions. The bold line is a first order microcanonical
phase transition line (between gaseous and core-halo phases), ending
at a \textbf{microcanonical critical point (Mcp)}. The grey line is a
canonical first order transition line (convexification singularity),
with a \textbf{canonical critical point (Ccp)}. The dashed grey
line is a line of canonical spinodal point. It crosses the
microcanonical first order phase transition at~Cr. The hashed zone is
the inequivalence area, absent in the canonical ensemble; the
doubly hashed is the negative specific region. The bold dashed line is a line of microcanonical stability change,
marking the limits of stability of the gaseous and core-halo phases
(such microcanonical stability changes are not described in our 
classification).  For \protect\( 1/\mu =0\protect \),
we recover the self-gravitating isothermal collapse (CE) and gravitational
phase transition (MCE) points. A similar phase
diagram has been done independently in \cite{Chavanis_Rieutord}.\\

\begin{figure}[p]
{\centering \resizebox*{\textwidth}{!}{\includegraphics
{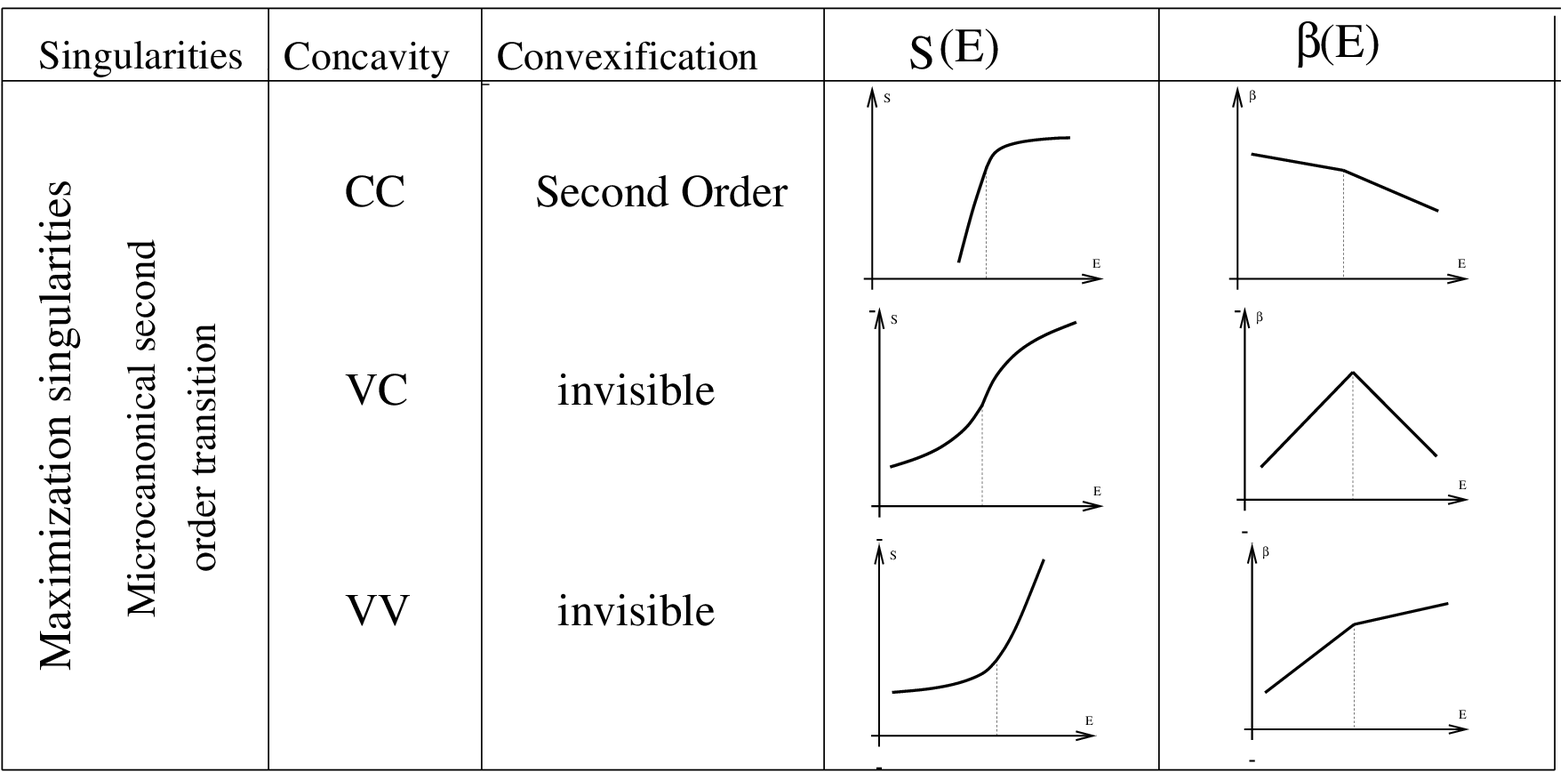}} \par}
\caption{\label{codim0_Symmetrie} New codimension 0 singularities 
for systems with symmetry.}
\vskip 3cm
\end{figure}

\noindent
Fig.~\ref{codim0_Symmetrie}: New codimension 0 singularities 
for systems with symmetry.\\

\begin{figure}[p]
{\centering \resizebox*{\textwidth}{!}
{\includegraphics{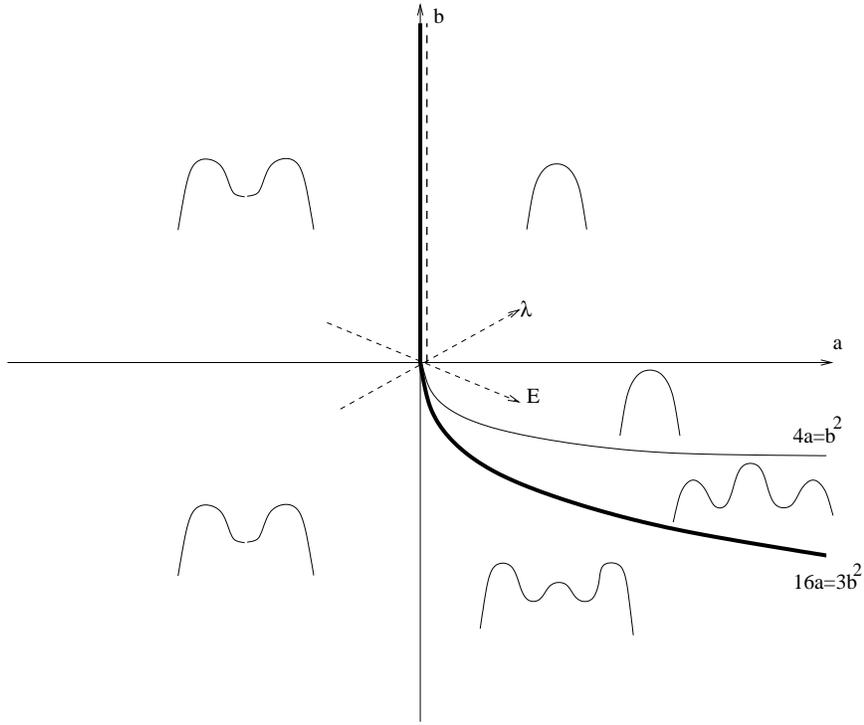}} \par}
\caption{\label{Fig_tricritique} Transition lines in the vicinity of a 
  \textbf{microcanonical tricritical point}, from the normal form
  \protect\( S\left( E,\lambda \right) =\max _{m}\left\{
    -m^{6}-3bm^{4}/2-3am^{2}\right\} \protect \). The tricritical
  point is reached for $a=b=0$. Small insets show the typical behavior
  of \protect\( s_{a,b}\left( m\right) \protect \) in the various
  areas. The curve \protect\( 4a=b^{2}\protect \) corresponds to the
  appearance of three local maxima. The bold curve (\protect\(
  16a=3b^{2},\, b<0\protect \)) is a first order transition line. The
  bold-dashed curve is a second order transition line, with negative
  concavity jump when going on the dashed side of the curve.
  \protect\( a\protect \) and \protect\( b\protect \) are linear
  combinations of \protect\( E\protect \) and \protect\( \lambda
  \protect \). Some possible directions for \protect\( E\protect \)
  and \protect\( \lambda \protect \) are represented by the dashed
  arrows; with these orientations, the transition is first order for 
  $\lambda<0$ and second order for $\lambda>0$.} \vskip 5cm
\end{figure}

\noindent
Fig.~\ref{Fig_tricritique}: Transition lines in the vicinity of a 
  \textbf{microcanonical tricritical point}, from the normal form
  \protect\( S\left( E,\lambda \right) =\max _{m}\left\{
    -m^{6}-3bm^{4}/2-3am^{2}\right\} \protect \). The tricritical
  point is reached for $a=b=0$. Small insets show the typical behavior
  of \protect\( s_{a,b}\left( m\right) \protect \) in the various
  areas. The curve \protect\( 4a=b^{2}\protect \) corresponds to the
  appearance of three local maxima. The bold curve (\protect\(
  16a=3b^{2},\, b<0\protect \)) is a first order transition line. The
  bold-dashed curve is a second order transition line, with negative
  concavity jump when going on the dashed side of the curve.
  \protect\( a\protect \) and \protect\( b\protect \) are linear
  combinations of \protect\( E\protect \) and \protect\( \lambda
  \protect \). Some possible directions for \protect\( E\protect \)
  and \protect\( \lambda \protect \) are represented by the dashed
  arrows; with these orientations, the transition is first order for 
  $\lambda<0$ and second order for $\lambda>0$.\\

\begin{figure}[p]
{\centering \resizebox*{\textwidth}{!}
{\includegraphics{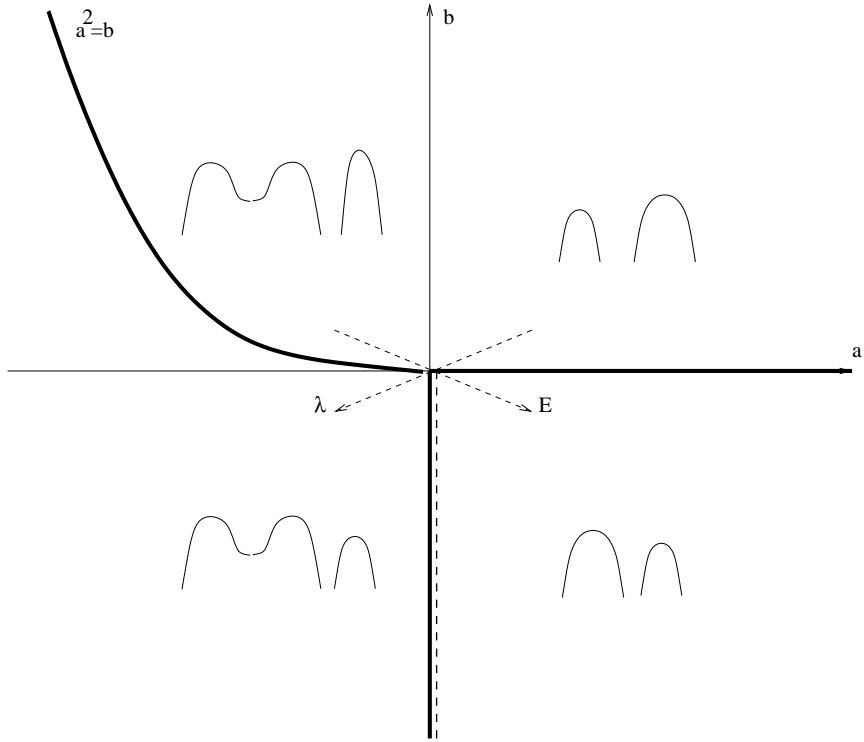}} \par}
\caption{\label{Fig_crossing_order1_order2}  Transition lines in the
  vicinity of a
  \textbf{microcanonical critical end point}, from the normal form
  \protect\( S\left( E,\lambda \right) =\max _{m}\left\{
    -m^{4}-2am^{2}\, ;\, -m^{2}+b\right\} \protect \). Small insets
  show the typical behavior of \protect\( s_{a,b}\left( m\right)
  \protect \) in the various areas, the lhs figure referring to the
  quartic part. Meaning of curves and arrows are as in
  Fig.~\ref{Fig_tricritique}.}
 \vskip 8cm
\end{figure}

\noindent
Fig.~\ref{Fig_crossing_order1_order2}: Transition lines in the
  vicinity of a
  \textbf{microcanonical critical end point}, from the normal form
  \protect\( S\left( E,\lambda \right) =\max _{m}\left\{
    -m^{4}-2am^{2}\, ;\, -m^{2}+b\right\} \protect \). Small insets
  show the typical behavior of \protect\( s_{a,b}\left( m\right)
  \protect \) in the various areas, the lhs figure referring to the
  quartic part. Meaning of curves and arrows are as in
  Fig.~\ref{Fig_tricritique}.\\

\begin{figure}[p]

{\centering \resizebox*{\textwidth}{!}
{\includegraphics{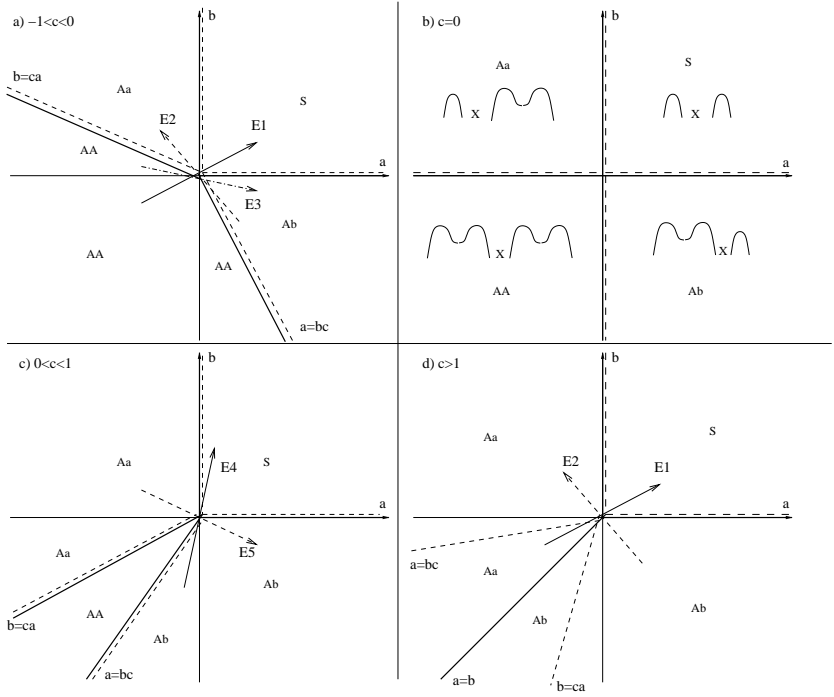}} \par}
\caption{\label{Fig_crossing_order2_order2} Transition lines in the
  vicinity of a crossing of two microcanonical second order
  transitions, from the normal form \protect\( S\left( E,\lambda
  \right) =\max _{m_{1},m_{2}} \left\{
    -m_{1}^{4}-2am_{1}^{2}-m_{2}^{4}-2bm_{2}^{2}-2cm^{2}_{1}m^{2}_{2}
  \right\} \protect \) (see text). Letters S, Aa, Ab and AA refer to
  the four phases involved: Symmetric S \protect\(
  m_{1}=m_{2}=0\protect \), Antisymmetric Ab: \protect\( m_{1}=0\, \,
  \textrm{and}\, \, m_{2}\neq 0\protect \), Antisymmetric Aa:
  \protect\( m_{1}\neq 0\, \, \textrm{and}\, \, m_{2}=0\protect \),
  and Antisymmetric AA: \protect\( m_{1}\neq 0\, \, \textrm{and}\, \,
  m_{2}\neq 0\protect \). Their entropy is respectively \protect\(
  S_{S}=0 \protect \), \protect\( S_{Aa}=a^{2}\protect \), \protect\(
  S_{Ab}=b^{2}\protect \) and \protect\( S_{AA}=\left(
    b^{2}+a^{2}-2cab\right) /\left( 1-c^{2}\right) \protect \).
  Panels a, b and c represent respectively \protect\( -1<c<0\protect
  \), \protect\( c=0\protect \), \protect\( 0<c<1\protect \) and all
  correspond to a {\bf second order quadruple point} (this is not a
  standard term). Panel d is for \protect\( c>1\protect \) and
  corresponds to a {\bf microcanonical bicritical point}.  For
  \protect\( c=0\protect \), small insets show the typical behavior of
  \protect\( s_{a,b}\left( m_{1},m_{2}\right) \protect \) in the
  various areas, the rhs (resp. lhs) figure referring to the
  dependence on \protect\( m_{1}\protect \) (resp \protect\(
  m_{2}\protect \)). The second order lines are represented by
  bold-dashed curve, with negative concavity jump when going on the
  dashed side of the curve. Possible directions for \protect\(
  E\protect \) are represented by the arrows in panels a), c) and d).
  } \vskip 3cm
\end{figure}

\noindent
Fig.~\ref{Fig_crossing_order2_order2}: Transition lines in the
  vicinity of a crossing of two microcanonical second order
  transitions, from the normal form\\ 
\protect\( S\left( E,\lambda
\right) =\max _{m_{1},m_{2}} \left\{
  -m_{1}^{4}-2am_{1}^{2}-m_{2}^{4}-2bm_{2}^{2}-2cm^{2}_{1}m^{2}_{2}
\right\} \protect \) (see text).\\ 
Letters S, Aa, Ab and AA refer to the
four phases involved: Symmetric S \protect\( m_{1}=m_{2}=0\protect \),
Antisymmetric Ab: \protect\( m_{1}=0\, \, \textrm{and}\, \, m_{2}\neq
0\protect \), Antisymmetric Aa: \protect\( m_{1}\neq 0\, \,
\textrm{and}\, \, m_{2}=0\protect \), and Antisymmetric AA: \protect\(
m_{1}\neq 0\, \, \textrm{and}\, \, m_{2}\neq 0\protect \). Their
entropy is respectively \protect\( S_{S}=0 \protect \), \protect\(
S_{Aa}=a^{2}\protect \), \protect\( S_{Ab}=b^{2}\protect \) and
\protect\( S_{AA}=\left( b^{2}+a^{2}-2cab\right) /\left(
  1-c^{2}\right) \protect \).  Panels a, b and c represent
respectively \protect\( -1<c<0\protect \), \protect\( c=0\protect \),
\protect\( 0<c<1\protect \) and all correspond to a {\bf second order
  quadruple point} (this is not a standard term). Panel d is for
\protect\( c>1\protect \) and corresponds to a {\bf microcanonical
  bicritical point}.  For \protect\( c=0\protect \), small insets show
the typical behavior of \protect\( s_{a,b}\left( m_{1},m_{2}\right)
\protect \) in the various areas, the rhs (resp. lhs) figure referring
to the dependence on \protect\( m_{1}\protect \) (resp \protect\(
m_{2}\protect \)). The second order lines are represented by
bold-dashed curve, with negative concavity jump when going on the
dashed side of the curve. Possible directions for \protect\( E\protect
\) are represented by the arrows in panels a), c) and d).\\

\begin{figure}[p]
{\centering \resizebox*{\textwidth}{!}
{\includegraphics{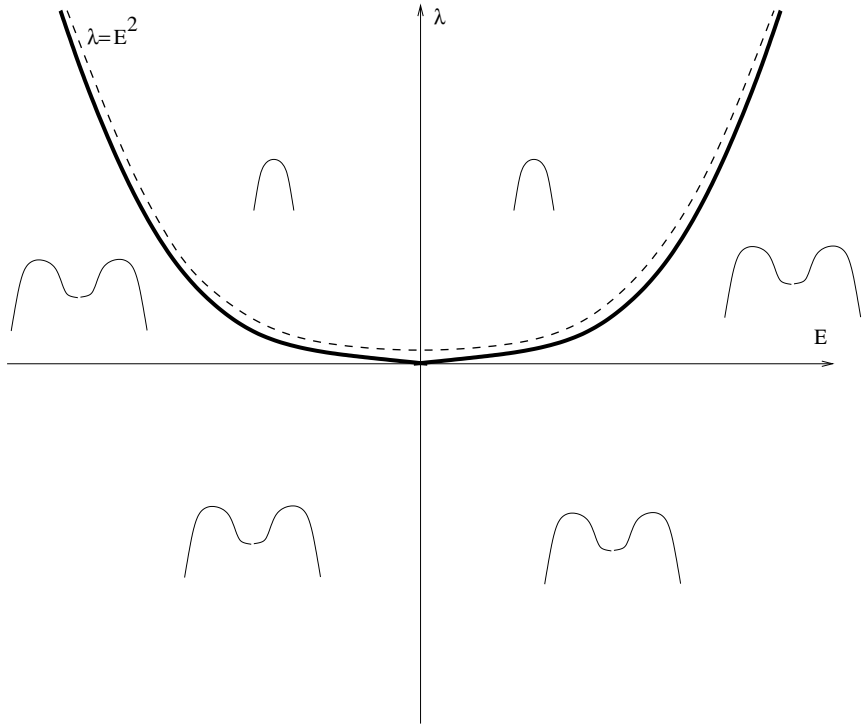}} \par}
\caption{\label{Fig_second_order_azeotropy}  Transition lines in the
  vicinity of a \textbf{second order
azeotropy}, from the normal form \protect\( S\left( E,\lambda \right) =\max _{m}\left\{ -m^{4}+2\left( E^{2}-\lambda \right) m^{2}\right\} \protect \).
The bold-dashed curve is a second order transition line, with negative
concavity jump when going on the dashed side of the curve. } \vskip
5cm
\end{figure}

\noindent
Fig.~\ref{Fig_second_order_azeotropy}: Transition lines in the
  vicinity of a \textbf{second order
azeotropy}, from the normal form \protect\( S\left( E,\lambda \right) =\max _{m}\left\{ -m^{4}+2\left( E^{2}-\lambda \right) m^{2}\right\} \protect \).
The bold-dashed curve is a second order transition line, with negative
concavity jump when going on the dashed side of the curve.\\

\begin{figure}[p]
{\centering \resizebox*{\textwidth}{!}
{\includegraphics{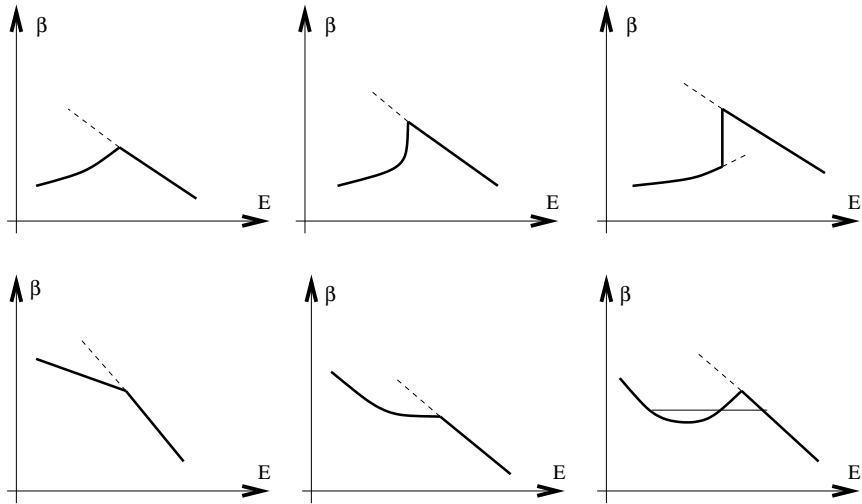}} \par}
\caption{ \label{Fig_Tricritique2} 
  Schematic caloric curves around a microcanonical VC (upper row) and
  a canonical (lower row) tricritical point, respectively before
  (left), at (center), and after (right) the singularity. The bold
  lines are the microcanonical solution as well as the canonical one
  when they are equivalent; the thin line is the canonical solution in
  the inequivalence range (not drawn on the top row); the dashed lines
  represent microcanonically metastable or unstable
  branches. Note the ensemble inequivalence onset around a canonical
  tricritical point.}
\end{figure}

\noindent
Fig.~\ref{Fig_Tricritique2}: Schematic caloric curves around a
microcanonical VC (upper row) and a canonical (lower row) tricritical
point, respectively before (left), at (center), and after (right) the
singularity. The bold lines is the microcanonical solution as well as
the canonical one when they are equivalent; the thin line is the
canonical solution in the inequivalence range (not drawn on the top
row); the dashed line represents microcanonically
metastable or unstable branches. Note the ensemble inequivalence onset
around a canonical tricritical
point.\\

\begin{figure}[p]
{\centering \resizebox*{\textwidth}{!}
{\includegraphics{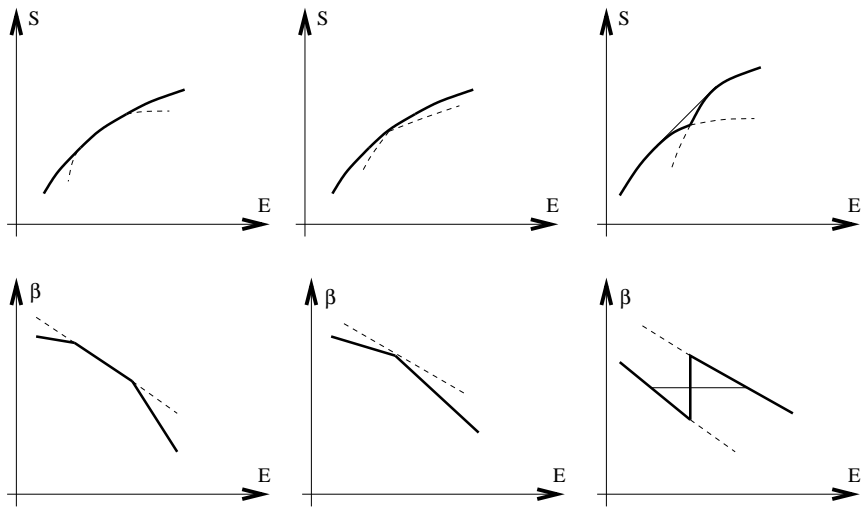}} \par}
\caption{ \label{Fig_Bicritique} Schematic entropic and caloric curves
  around a bicritical point, of CCC$\to$CC type, before, at and
  after the singularity. Lines have the same signification as
  in Fig.~\ref{Fig_Tricritique2}. Note the other type of
  inequivalence onset.}
\end{figure}

\noindent
Fig.~\ref{Fig_Bicritique}: Schematic entropic and caloric curves
around a bicritical point, of CCC$\to$CC type, before, at and after
the singularity. Lines have the same signification as in
Fig.~\ref{Fig_Tricritique2}. Note the other type of inequivalence
onset.\\

\noindent
Table.~\ref{Fig_codim1_sym}: This table summarizes the new codimension
1 singularities, when some symmetry of the system is considered. The
third column is the number of singularities of a given type (the second number when
convexity properties are taken into account, the first one when not). The fourth column is
the number of singularities visible in the canonical ensemble together
with their type. The last column gives the number of singularities
associated with the appearance/disappearance of an energy range of
ensemble inequivalence. For instance, they are 4 types of microcanonical
second order quadruple points, 20 when convexity is taken into account, 6 of which are canonically visible,
giving respectively 4 canonical second order quadruple points and 2 canonical 
bicritical points. This last one is associated to the onset 
of ensemble inequivalence.

\begin{table}[p]
\vskip 1cm
\caption{\label{Fig_codim1_sym} This table summarizes the new codimension
1 singularities, when some symmetry of the system is considered. The
third column is the number of singularities of a given type: the first
number takes into account only the number of possible ways a
singularity can be seen on the entropic curve; the second number takes 
into account in addition the different convexity
configurations. The fourth column is the number of singularities
visible in the canonical ensemble together with their type. The last
column gives the number of singularities associated with the
appearance/disappearance of an energy range of ensemble
inequivalence. For instance, there are 4 types of microcanonical second
order quadruple points, 20 when convexity is taken into account, 6 of
which are canonically visible, giving respectively the four types of 
canonical second order quadruple points and two canonical bicritical
points. These last ones are associated with ensemble inequivalence
onset.}

{\centering {\footnotesize \begin{tabular}{|c|c|c|c|c|}
\hline 
\begin{tabular}{c}
Singularity\\
type
\end{tabular}&
\begin{tabular}{c}
Microcanonical\\ 
singularity
\end{tabular}&
Nbr&
Canonical (Nbr)&
Ineq.\\
\hline
\hline 
Maximization&
{\footnotesize \begin{tabular}{c}
Tricritical\\
Critical end point\\
2nd order quadruple\\
\\
Bicritical\\
2nd order azeotropy\\
\end{tabular}}&
\begin{tabular}{c}
1/2\\
1/6\\
4/20\\
\\
2/8\\
1/2\\
\end{tabular}&
\begin{tabular}{c}
None\\
None\\
2nd order quadruple (4)\\
Bicritical (2)\\
Bicritical (2)\\
2nd order azeotropy (1)\\
\end{tabular}&
\begin{tabular}{c}
None\\
None\\
None\\
2\\
2\\
None\\
\end{tabular}\\
\hline 
Convexity&
\begin{tabular}{c}
2nd order\\  
+inflexion
\end{tabular}&
1/2&
Tricritical (1)&
1\\
\hline 
Convexification&
\begin{tabular}{c}
2nd order\\
+convexification
\end{tabular}&
1&
Critical end point (1)&
none\\
\hline
\end{tabular}}\footnotesize \par}
\end{table}

\end{document}